\documentclass[twocolumn,prd,amsmath,amssymb,superscriptaddress,reprint]{revtex4}

\usepackage{graphicx}
\usepackage{dcolumn}
\usepackage{bm}
\usepackage{color}
 \graphicspath{{./figures/}} 
 \DeclareGraphicsExtensions{.pdf,.jpg,.jpeg}  
 \usepackage{multirow}
 \usepackage{placeins}
\usepackage{hyperref}
\usepackage{subfig}

\def\met{\ensuremath{\not\!\!{E_\textrm{T}}}}
\def\vecmet{\ensuremath{\not\!\! \vec{E}_\textrm{T}}}

\begin{document}

\title{\boldmath Measurement of the differential cross sections for $W$-boson production in association with jets in $p\bar{p}$ collisions at $\sqrt{s}=1.96$~TeV}

\affiliation{Institute of Physics, Academia Sinica, Taipei, Taiwan 11529, Republic of China}
\affiliation{Argonne National Laboratory, Argonne, Illinois 60439, USA}
\affiliation{University of Athens, 157 71 Athens, Greece}
\affiliation{Institut de Fisica d'Altes Energies, ICREA, Universitat Autonoma de Barcelona, E-08193, Bellaterra (Barcelona), Spain}
\affiliation{Baylor University, Waco, Texas 76798, USA}
\affiliation{Istituto Nazionale di Fisica Nucleare Bologna, \ensuremath{^{kk}}University of Bologna, I-40127 Bologna, Italy}
\affiliation{University of California, Davis, Davis, California 95616, USA}
\affiliation{University of California, Los Angeles, Los Angeles, California 90024, USA}
\affiliation{Instituto de Fisica de Cantabria, CSIC-University of Cantabria, 39005 Santander, Spain}
\affiliation{Carnegie Mellon University, Pittsburgh, Pennsylvania 15213, USA}
\affiliation{Enrico Fermi Institute, University of Chicago, Chicago, Illinois 60637, USA}
\affiliation{Comenius University, 842 48 Bratislava, Slovakia; Institute of Experimental Physics, 040 01 Kosice, Slovakia}
\affiliation{Joint Institute for Nuclear Research, RU-141980 Dubna, Russia}
\affiliation{Duke University, Durham, North Carolina 27708, USA}
\affiliation{Fermi National Accelerator Laboratory, Batavia, Illinois 60510, USA}
\affiliation{University of Florida, Gainesville, Florida 32611, USA}
\affiliation{Laboratori Nazionali di Frascati, Istituto Nazionale di Fisica Nucleare, I-00044 Frascati, Italy}
\affiliation{University of Geneva, CH-1211 Geneva 4, Switzerland}
\affiliation{Glasgow University, Glasgow G12 8QQ, United Kingdom}
\affiliation{Harvard University, Cambridge, Massachusetts 02138, USA}
\affiliation{Division of High Energy Physics, Department of Physics, University of Helsinki, FIN-00014, Helsinki, Finland; Helsinki Institute of Physics, FIN-00014, Helsinki, Finland}
\affiliation{University of Illinois, Urbana, Illinois 61801, USA}
\affiliation{The Johns Hopkins University, Baltimore, Maryland 21218, USA}
\affiliation{Institut f\"{u}r Experimentelle Kernphysik, Karlsruhe Institute of Technology, D-76131 Karlsruhe, Germany}
\affiliation{Center for High Energy Physics: Kyungpook National University, Daegu 702-701, Korea; Seoul National University, Seoul 151-742, Korea; Sungkyunkwan University, Suwon 440-746, Korea; Korea Institute of Science and Technology Information, Daejeon 305-806, Korea; Chonnam National University, Gwangju 500-757, Korea; Chonbuk National University, Jeonju 561-756, Korea; Ewha Womans University, Seoul, 120-750, Korea}
\affiliation{Ernest Orlando Lawrence Berkeley National Laboratory, Berkeley, California 94720, USA}
\affiliation{University of Liverpool, Liverpool L69 7ZE, United Kingdom}
\affiliation{University College London, London WC1E 6BT, United Kingdom}
\affiliation{Centro de Investigaciones Energeticas Medioambientales y Tecnologicas, E-28040 Madrid, Spain}
\affiliation{Massachusetts Institute of Technology, Cambridge, Massachusetts 02139, USA}
\affiliation{University of Michigan, Ann Arbor, Michigan 48109, USA}
\affiliation{Michigan State University, East Lansing, Michigan 48824, USA}
\affiliation{Institution for Theoretical and Experimental Physics, ITEP, Moscow 117259, Russia}
\affiliation{University of New Mexico, Albuquerque, New Mexico 87131, USA}
\affiliation{The Ohio State University, Columbus, Ohio 43210, USA}
\affiliation{Okayama University, Okayama 700-8530, Japan}
\affiliation{Osaka City University, Osaka 558-8585, Japan}
\affiliation{University of Oxford, Oxford OX1 3RH, United Kingdom}
\affiliation{Istituto Nazionale di Fisica Nucleare, Sezione di Padova, \ensuremath{^{ll}}University of Padova, I-35131 Padova, Italy}
\affiliation{University of Pennsylvania, Philadelphia, Pennsylvania 19104, USA}
\affiliation{Istituto Nazionale di Fisica Nucleare Pisa, \ensuremath{^{mm}}University of Pisa, \ensuremath{^{nn}}University of Siena, \ensuremath{^{oo}}Scuola Normale Superiore, I-56127 Pisa, Italy, \ensuremath{^{pp}}INFN Pavia, I-27100 Pavia, Italy, \ensuremath{^{qq}}University of Pavia, I-27100 Pavia, Italy}
\affiliation{University of Pittsburgh, Pittsburgh, Pennsylvania 15260, USA}
\affiliation{Purdue University, West Lafayette, Indiana 47907, USA}
\affiliation{University of Rochester, Rochester, New York 14627, USA}
\affiliation{The Rockefeller University, New York, New York 10065, USA}
\affiliation{Istituto Nazionale di Fisica Nucleare, Sezione di Roma 1, \ensuremath{^{rr}}Sapienza Universit\`{a} di Roma, I-00185 Roma, Italy}
\affiliation{Mitchell Institute for Fundamental Physics and Astronomy, Texas A\&M University, College Station, Texas 77843, USA}
\affiliation{Istituto Nazionale di Fisica Nucleare Trieste, \ensuremath{^{ss}}Gruppo Collegato di Udine, \ensuremath{^{tt}}University of Udine, I-33100 Udine, Italy, \ensuremath{^{uu}}University of Trieste, I-34127 Trieste, Italy}
\affiliation{University of Tsukuba, Tsukuba, Ibaraki 305, Japan}
\affiliation{Tufts University, Medford, Massachusetts 02155, USA}
\affiliation{Waseda University, Tokyo 169, Japan}
\affiliation{Wayne State University, Detroit, Michigan 48201, USA}
\affiliation{University of Wisconsin-Madison, Madison, Wisconsin 53706, USA}
\affiliation{Yale University, New Haven, Connecticut 06520, USA}

\author{T.~Aaltonen}
\affiliation{Division of High Energy Physics, Department of Physics, University of Helsinki, FIN-00014, Helsinki, Finland; Helsinki Institute of Physics, FIN-00014, Helsinki, Finland}
\author{S.~Amerio\ensuremath{^{ll}}}
\affiliation{Istituto Nazionale di Fisica Nucleare, Sezione di Padova, \ensuremath{^{ll}}University of Padova, I-35131 Padova, Italy}
\author{D.~Amidei}
\affiliation{University of Michigan, Ann Arbor, Michigan 48109, USA}
\author{A.~Anastassov\ensuremath{^{w}}}
\affiliation{Fermi National Accelerator Laboratory, Batavia, Illinois 60510, USA}
\author{A.~Annovi}
\affiliation{Laboratori Nazionali di Frascati, Istituto Nazionale di Fisica Nucleare, I-00044 Frascati, Italy}
\author{J.~Antos}
\affiliation{Comenius University, 842 48 Bratislava, Slovakia; Institute of Experimental Physics, 040 01 Kosice, Slovakia}
\author{G.~Apollinari}
\affiliation{Fermi National Accelerator Laboratory, Batavia, Illinois 60510, USA}
\author{J.A.~Appel}
\affiliation{Fermi National Accelerator Laboratory, Batavia, Illinois 60510, USA}
\author{T.~Arisawa}
\affiliation{Waseda University, Tokyo 169, Japan}
\author{A.~Artikov}
\affiliation{Joint Institute for Nuclear Research, RU-141980 Dubna, Russia}
% ...

\collaboration{CDF Collaboration}
\altaffiliation[With visitors from]{
\ensuremath{^{a}}University of British Columbia, Vancouver, BC V6T 1Z1, Canada,
\ensuremath{^{b}}Istituto Nazionale di Fisica Nucleare, Sezione di Cagliari, 09042 Monserrato (Cagliari), Italy,
\ensuremath{^{c}}University of California Irvine, Irvine, CA 92697, USA,
\ensuremath{^{d}}Institute of Physics, Academy of Sciences of the Czech Republic, 182~21, Czech Republic,
\ensuremath{^{e}}CERN, CH-1211 Geneva, Switzerland,
\ensuremath{^{f}}Cornell University, Ithaca, NY 14853, USA,
\ensuremath{^{g}}University of Cyprus, Nicosia CY-1678, Cyprus,
\ensuremath{^{h}}Office of Science, U.S. Department of Energy, Washington, DC 20585, USA,
\ensuremath{^{i}}University College Dublin, Dublin 4, Ireland,
\ensuremath{^{j}}ETH, 8092 Z\"{u}rich, Switzerland,
\ensuremath{^{k}}University of Fukui, Fukui City, Fukui Prefecture, Japan 910-0017,
\ensuremath{^{l}}Universidad Iberoamericana, Lomas de Santa Fe, M\'{e}xico, C.P. 01219, Distrito Federal,
\ensuremath{^{m}}University of Iowa, Iowa City, IA 52242, USA,
\ensuremath{^{n}}Kinki University, Higashi-Osaka City, Japan 577-8502,
\ensuremath{^{o}}Kansas State University, Manhattan, KS 66506, USA,
\ensuremath{^{p}}Brookhaven National Laboratory, Upton, NY 11973, USA,
\ensuremath{^{q}}Istituto Nazionale di Fisica Nucleare, Sezione di Lecce, Via Arnesano, I-73100 Lecce, Italy,
\ensuremath{^{r}}Queen Mary, University of London, London, E1 4NS, United Kingdom,
\ensuremath{^{s}}University of Melbourne, Victoria 3010, Australia,
\ensuremath{^{t}}Muons, Inc., Batavia, IL 60510, USA,
\ensuremath{^{u}}Nagasaki Institute of Applied Science, Nagasaki 851-0193, Japan,
\ensuremath{^{v}}National Research Nuclear University, Moscow 115409, Russia,
\ensuremath{^{w}}Northwestern University, Evanston, IL 60208, USA,
\ensuremath{^{x}}University of Notre Dame, Notre Dame, IN 46556, USA,
\ensuremath{^{y}}Universidad de Oviedo, E-33007 Oviedo, Spain,
\ensuremath{^{z}}CNRS-IN2P3, Paris, F-75205 France,
\ensuremath{^{aa}}Universidad Tecnica Federico Santa Maria, 110v Valparaiso, Chile,
\ensuremath{^{bb}}Sejong University, Seoul 143-747, Korea,
\ensuremath{^{cc}}The University of Jordan, Amman 11942, Jordan,
\ensuremath{^{dd}}Universite catholique de Louvain, 1348 Louvain-La-Neuve, Belgium,
\ensuremath{^{ee}}University of Z\"{u}rich, 8006 Z\"{u}rich, Switzerland,
\ensuremath{^{ff}}Massachusetts General Hospital, Boston, MA 02114 USA,
\ensuremath{^{gg}}Harvard Medical School, Boston, MA 02114 USA,
\ensuremath{^{hh}}Hampton University, Hampton, VA 23668, USA,
\ensuremath{^{ii}}Los Alamos National Laboratory, Los Alamos, NM 87544, USA,
\ensuremath{^{jj}}Universit\`{a} degli Studi di Napoli Federico II, I-80138 Napoli, Italy
}
\noaffiliation

\date{\today}

\begin{abstract}

This paper presents a study of the production of a single $W$ boson in association with one or more jets in proton-antiproton collisions at $\sqrt{s}=1.96$\,TeV, using the entire data set collected in 2001-2011 by the Collider Detector at Fermilab at the Tevatron, which corresponds to an integrated luminosity of $9.0$\,fb$^{-1}$. The $W$ boson is identified through its leptonic decays into electron and muon.
The production cross sections are measured for each leptonic decay mode and combined after testing that the ratio of the $W(\rightarrow \mu\nu)+$jets cross section to the $W(\rightarrow e\nu)+$jets cross section agrees with the hypothesis of $e$-$\mu$ lepton universality. 
The combination of measured cross sections, differential in the inclusive jet multiplicity ($W+\geqslant N$ jets with $N=1,\,2,\,3, \textrm{or}\, 4$) and in the transverse energy of the leading jet, are compared with theoretical predictions.

\end{abstract}

\maketitle

\section{\label{sec:I} Introduction}
The production of $W$ bosons in association with jets in proton-antiproton ($p\bar{p}$) collisions requires high momentum transfer between the interacting constituents of the incoming hadrons. It follows that this process is suitable for testing perturbative quantum chromodynamics (pQCD). Moreover, it is an important background for standard-model (SM) processes such as Higgs-boson and $t\bar{t}$ production at hadron colliders. 
This paper reports measurements of the inclusive production cross sections $\sigma_{N}= \sigma(W(\rightarrow \ell \nu) + \geqslant N$ jets), where $\ell$ is either an electron or a muon, for each of the jet multiplicities $N=1,\,2,\,3, \textrm{or}\, 4$  in $p\bar{p}$-collisions. In addition to these inclusive cross sections, differential cross sections ($d\sigma_{1}/dE^{\rm jet}_\textrm{T}$) as functions of the leading-jet energy transverse to the beam direction  ($E^{\rm jet}_\textrm{T}$) are presented.  

These measurements are performed by selecting $W$-boson decays with one electron or one muon detected in the central region of the Collider Detector at Fermilab (CDF) and by requiring the presence of at least one hadronic jet. The transverse energies (momenta)~\cite{transverse} of electrons (muons) are required to exceed 25\,GeV (25\,GeV/$c$) as are the transverse energies of jets. Jets are defined using a cone-based jet clustering algorithm.
Although the presence of one high-transverse-momentum lepton is a distinctive signature for identifying the $W$ boson, background contamination remains significant. One of the
most challenging tasks is the subtraction from the selected sample of the multijet background made of jets that have experimental signatures similar to those of the leptons and are therefore reconstructed as electrons or as muons. The techniques used to model this background are optimized to obtain a better identification of the signal sample and to reduce the systematic uncertainties of the results. The measured cross sections are then corrected for detector effects using an unfolding procedure for a straightforward comparison with theoretical predictions at the particle level.

The measurements are obtained using the entire $p\bar{p}$ collision data set collected with CDF~II detector in Run~II (2001-2011) at the Tevatron collider, corresponding to $9.0$\,fb$^{-1}$ of integrated luminosity. They follow previous studies of jet pairs produced in association with a $W$ boson~\cite{PRD89092001(2014)} and a measurement of $W+$jets production cross sections that considered only  electron final states in a sample corresponding to 320\,pb$^{-1}$ of integrated luminosity~\cite{PRD77011108(2008)}. The current measurement improves upon previous CDF studies in that it uses the entire Run II data set and it includes the investigation of the muon channel, resulting in more data and a partially complementary set of systematic uncertainties.  
Recent studies of the $W+$jets process in $p\bar{p}$ collisions have been reported by the D0 \cite{PLB705200(2011)} collaboration and in $pp$ collisions by the ATLAS~\cite{EPJC7582(2015)}, CMS~\cite{PLB74112(2015)}, and LHCb~\cite{JHEP052016131} collaborations.

This paper is structured as follows. The CDF II detector is described in Sec.\,\ref{sec:II}. In Sec.\,\ref{sec:III}  the details of the $W+$\,jets event selection are presented. Section\,\ref{sec:IV} describes how the background is estimated and subtracted. The procedure used to unfold the data is presented in Sec.\,\ref{sec:V}, and the systematic uncertainties are discussed in Sec.\,\ref{sec:VI}. Section\,\ref{sec:VII} describes the combination of electron and muon results. Comparisons of theoretical predictions with the data are discussed in Sec.\,\ref{sec:VIII}, and are presented with the results in Sec.\,\ref{sec:IX}. Finally, the results are summarized and conclusions are drawn in Sec.\,\ref{sec:X}. Appendices~\ref{APPI} and~\ref{APPII} detail the background validation and the unfolding procedure, respectively.

\section{\label{sec:II} The CDF II Detector}

The CDF II detector was a general-purpose apparatus that collected $p\bar{p}$ collision data from the Tevatron between 2001 and 2011~\cite{JPG342457(2007)}. The detector consisted of a tracking system contained in a 1.4\,T solenoidal magnetic field, surrounded by sampling calorimeters and muon detectors. 

The CDF~II detector was cylindrically symmetric around the beam axis. The coordinate system has its origin in the center of the detector, and consists of the radius $r$, the azimuthal angle $\phi$, and the polar angle $\theta$ measured from the $z$-axis, which is oriented along the incoming proton beam.
The pseudorapidity  is defined as $\eta = -\ln\left(\tan\frac{\theta}{2}\right)$; the transverse energy as $E_\textrm{T}=E \sin\theta$, $E$ being the energy detected by the calorimeters; and the transverse momentum as $p_\textrm{T}=p \sin\theta$, $p$ being the magnitude of the momentum reconstructed by the tracking system. The angular distance between two reconstructed particles or clusters of particles $P_{1}$ and $P_{2}$ is defined as $\Delta R\left(P_{1},P_{2}\right)=\sqrt{\left(\phi_{P_{1}}-\phi_{P_{2}}\right)^2 +\left(\eta_{P_{1}}-\eta_{P_{2}}\right)^2}$.

Charged particle  trajectories (tracks) were reconstructed by a silicon microstrip system~\cite{NIMA4471(2000),NIMA45384(2000)} located just outside the interaction region, surrounded by the central outer tracker~(COT)~\cite{NIMA526249(2004),NPPSB61230(1998)}. Together they provided high-resolution tracking information for pseudorapidities $|\eta|<1$. The silicon microstrip system consisted of a central part  (SVXII) which covered $|\eta| \leqslant 1$ and an intermediate part (ISL) which extended coverage (with degraded resolution) to $|\eta|= 2$. The SVXII comprised a layer of single-sided silicon microstrip detectors at 1.6 cm from the beam and a five-layer double-sided silicon microstrip detector at radii  ranging from 2.5 to 11~cm. The ISL was located between the radii of 19 and 29~cm at higher $|\eta|$. The transverse momentum resolution was $\sigma_{p_\textrm{T}}/p^2_\textrm{T}=0.0017\,($GeV$/c)^{-1}$.

The sampling calorimeter system was located outside the solenoid.  It included inner electromagnetic and outer hadronic calorimeters, both comprising central and forward (end-plug) sections.
The central section, which included the cylindrical central electromagnetic  (CEM)~\cite{NIMA267272(1988)} and central hadronic (CHA) calorimeters, followed by the hadronic end-wall  (WHA) calorimeter~\cite{NIMA267301(1988)} covered the range of pseudorapidy $|\eta|<1.1$. The end-plug electromagnetic (PEM) and hadronic (PHA) calorimeters extended the coverage up to $|\eta|<3.64$~\cite{NIMA453227(2000)}.  All calorimeter sections were  subdivided into projective modules (towers) pointing to the nominal interaction point. Each projective tower in the central region covered 0.1 in $\eta$ and 15$^\circ$ in $\phi$. The size of the  projective towers in the plug calorimeters changed progressively from 0.1 in $\eta$ and 7.5$^\circ$ in $\phi$ at $|\eta|=1.1$ to 0.5 in $\eta$ and 15$^\circ$ in $\phi$ at $|\eta|=3.64$~\cite{JPG342457(2007)}. Sampling of the energy deposited in all calorimeters was obtained by interleaving active scintillator with passive metal layers (lead in the electromagnetic and steel in the hadronic sections). Shower profiles were measured by strip detectors located near the shower maxima (at $\sim6$ radiation lengths) in the electromagnetic calorimeters: the central electromagnetic strip chambers (CES)  and the plug electromagnetic strip detector (PES)~\cite{NIMA412515(1998)} with  2~cm and 1.5~cm resolution, respectively. The unresolved gamma background is reduced  by  scintillator - tile preshower counters located near the front faces of all  electromagnetic calorimeters~\cite{NIMA431104(1999),NSSC}.

The muon detector~\cite{NIMA26833(1988),NIMA538358(2005)} included four independent detectors located behind the hadronic calorimeter. Coverage for pseudorapidities $|\eta|<0.6$ was provided by a central muon detector (CMU), located behind the central hadronic calorimeter and followed by a central muon upgrade (CMP) detector after an additional layer of shielding steel. The pseudorapidity region $0.6<|\eta|<1.0$ was covered by a central muon extension (CMX) detector. These three muon detectors comprised wire drift chambers operating in proportional mode interleaved with scintillator planes. Finally, coverage was extended to the region $1.0<|\eta|<1.5$ by the barrel muon upgrade (BMU) detector~\cite{BMU}.

Cherenkov counters located at small angle, $3.7<|\eta|<4.7$ were used to determine the luminosity by measuring the average rate of inelastic $p\bar{p}$ collisions in each bunch crossing~\cite{NIMA441366(2000)}.

\section{\label{sec:III} Event Reconstruction and Selection}

Events enriched in decays of a $W$ boson into an electron or muon, and a neutrino, are selected using an inclusive-lepton online event selection system (trigger)~\cite{JPG342457(2007)}.  

In the offline data reduction, electron and muon candidates are identified using standard requirements~\cite{JPG342457(2007)}.  Electron candidates are identified as charged particles whose trajectories geometrically match significant energy deposits in a few adjacent calorimeter cells, while muon candidates are tracks that match signals in the muon detectors and deposit no significant energy in the calorimeters. 
Electron and muon candidates are required to satisfy requirements on the minimum number of COT hits and the primary-vertex position~\cite{vertex}. Requirements are also applied to the fraction of particle energy, inferred from the momentum measurement, deposited in the calorimeter. At least 95\% of 
the energy is required in the electromagnetic calorimeter for the electrons, and little energy in both calorimeters for the muons. 
Selection requirements for the electron candidates include conditions, referred to as identification (ID) criteria~\cite{PRD89092001(2014)}, that are used to reduce 
the probability that a jet is misidentified as an electron. These ID criteria include requirements on the ratio between the energy deposited in the electromagnetic 
and hadronic calorimeters and on the shape and position of the shower produced by the electron candidate.  
Finally, both electron and muon candidates are required to be isolated~\cite{isolation}. Only the leptons in the central part of the detector ($|\eta|<1.0$), where the track reconstruction efficiency is optimal and the calorimeter is well instrumented, are considered.

Candidate $W+$jets events are selected from this inclusive lepton data set by requiring the presence of exactly one central electron or muon candidate with $E_\textrm{T}> 25$\;GeV or $p_\textrm{T}>25$\;GeV/$c$, respectively, and at least one jet. 

Jets are reconstructed using the JETCLU cone algorithm~\cite{PRD451448(1992)} with jet-radius parameter $R=\sqrt{(\Delta\phi)^2+(\Delta\eta)^2}=0.4$. Only jets having $E_\textrm{T}>25$\;GeV, $|\eta|<2$, electromagnetic fraction ({\it i.e.}, the fraction of the total calorimetric energy of the jet deposited in the electromagnetic calorimeter) lower than 0.9, and that are well-separated from the lepton candidate ($\Delta R(\ell, {\rm jets}) > 0.4$) are considered. The energy of each jet is corrected using the jet-energy scale (JES) correction detailed in Ref.~\cite{NIMA566375(2006)}. 

 A threshold is imposed on the transverse mass of the $W$ candidate~\cite{MTW}, $m_\textrm{T}^{W} > 40\;\textrm{GeV$/c\,^2$}$.

\section{\label{sec:IV} Background Modeling and Validation}

The resulting sample is expected to include two classes of backgrounds, i) electroweak and top-quark processes and ii) multijet production. Background of the first type is modeled by using simulated samples. A reliable model of the multijet background is particularly difficult to produce using simulation, so this background is estimated using data.

\subsection{Simulated background processes}

Electroweak (EW) processes consist of decays into electrons or muons of gauge bosons produced in $W(\rightarrow \tau\nu)$+\,jets, $Z (\rightarrow \ell^{+} \ell^{-})+$\,jets, and {\it WW}, {\it WZ}, and {\it ZZ} processes. Top-quark processes involve the production and decay of top quarks, singly or in pairs.
The two classes of processes are modeled using Monte Carlo (MC) samples. 
Samples of $Z+$\,jets and $W(\rightarrow \tau\nu)+$\,jets events are generated using \textsc{Alpgen v1.3}~\cite{alpgen} interfaced with \textsc{Pythia v6.3}~\cite{pythia} for parton showering and hadronization. The contribution of the underlying event~\cite{UE} is included in the \textsc{Pythia} generator using the Tevatron-tuned parameters of \textsc{tune a}~\cite{PRD89092001(2014)}, and final jets are matched to the original partons with the MLM matching procedure described in~\cite{alpgen}. Production of {\it WW}, {\it WZ}, and {\it ZZ} pairs and top-quark pairs is modeled with the \textsc{Pythia} event generator, also using \textsc{tune a} and assuming a top-quark mass of 172.5\;GeV/$c^2$. Production of single top quarks (both in  the $s$ channel and in the $t$ channel) is modeled with the \textsc{Madevent}~\cite{madevent} generator followed by \textsc{Pythia} for parton showering and hadronization~\cite{Aaltonen:2010jr}. All simulated samples are generated assuming the CTEQ5L parton distribution functions (PDFs)~\cite{Lai:1999wy}.

The contributions expected from each process are based on theoretical cross-section predictions. The rate of diboson production ({\it WW}, {\it WZ}, and {\it ZZ}\,) is scaled to the cross section calculated at next-to-leading order (NLO) in pQCD~\cite{PRD60113006(1999)}; the $t\bar{t}$ sample is normalized using a next-to-next-to-leading order plus next-to-next-to-leading logarithm (NNLO+NNLL) pQCD cross-section calculation~\cite{PRL110252004(2013)}; and the single top-quark process is normalized to approximate NNLO+NNLL calculations~\cite{PRD81054028(2010)} for the $s$ channel and approximate NNLO+NLL calculations~\cite{PRD83091503(2011)} for the $t$ channel. The $Z (\rightarrow \ell^{+} \ell^{-})+$\,jets and $W(\rightarrow \tau\nu)+$\,jets cross sections are normalized to leading order (LO) pQCD calculations~\cite{alpgen} and scaled by a $K$-factor of 1.4 to account for higher-order effects~\cite{PRL100102001(2008), PRD77011108(2008)}. The uncertainties of these cross sections are 3\% and 11\% for top-quark pair production and single-top-quark production, respectively; 20\% for $Z(\rightarrow \ell^{+} \ell^{-})+$\,jets and 40\% for $W(\rightarrow \tau\nu)+$jets; and a fully correlated 6\% for {\it WW}, {\it WZ}, and {\it ZZ}. The background contributions are also affected by the uncertainties in the integrated luminosity measurement (6\%)~\cite{NIMA49457(2002)}, the lepton acceptance (2.2\%) and the jet-energy scale (see Sec.~\ref{sec:VI} for details).

All generated samples are processed using the CDF II detector simulation based on \textsc{GEANT3}~\cite{GEANT3}, and the same event reconstruction and selection procedures used for the experimental data, described in Sec.~\ref{sec:III}, are applied. Moreover, the events in each simulated sample are weighted so that the distribution of the number of reconstructed primary vertices, due to the additional $p\bar{p}$ interactions in the same bunch crossing (pileup), matches the distribution in the data.

\subsection{Multijet background}

Multijet background events enter the signal sample if one of the jets is incorrectly identified as a lepton. This background gives a large contribution in the electron channel, but is almost negligible in the muon channel because a jet, in order to mimic a muon, must also generate a matching track in the muon detectors. The multijet background  is modeled using data, following Ref.\,\cite{PRD89092001(2014)}. The background data samples are obtained from the same data set as that used for the analysis (and described in Sec.\,\ref{sec:III}) by requiring the failure of two (one) of the electron (muon) ID criteria~\cite{JPG342457(2007)}. 

In the electron channel, the multijet background events modeled in this way are referred to as ``nonelectron'' events~\cite{PRD89092001(2014)}. Only the ID criteria that introduce the least bias in the kinematic distributions of nonelectrons ($i.e.$, the fraction of energy deposited in the hadronic calorimeter by the electron candidate and the shape its shower) are inverted, so as to minimize differences with respect to the candidate electrons.

 The $E_\textrm{T}$ of nonelectron in the $W+2$\,jet sample is tuned following Ref.\,\cite{PRD89092001(2014)} and  the tuning procedure is generalized to other jet multiplicities. The tuning procedure includes two steps.
First, the contamination from all processes with a real lepton ($e.g.$, weak-boson decay) is evaluated using a MC technique and subtracted from the nonelectron event sample as a function of the variable of interest. Then, in order to model the kinematic properties of the event correctly, the $E_\textrm{T}$ of the nonelectron is taken to be the $E_\textrm{T}$ of the corresponding jet ({\it i.e.}, the jet with $\Delta R<0.4$ with respect to the nonelectron). After this procedure, the following two corrections to the $E_\textrm{T}$ of the jet producing a nonelectron have been applied.

The first correction, called the ``nonelectron energy-scale correction'', accounts for the difference in the energy scale between a jet producing a nonelectron and a jet producing a misidentified electron, {\it i.e.}, a jet fulfilling all the electron selection criteria. To correct the nonelectron transverse energy, the same energy-scale correction as was previously evaluated for the $W + 2$\;jets sample~\cite{PRD89092001(2014)} is used. This $E_\textrm{T}$ correction is tested in a multijet-enriched region (control region, CR) and shows very good agreement between data and MC expectations. The CR is defined by reversing the $W$-boson transverse-mass requirement for the signal region (SR), {\it i.e.}, by requiring that $m^W_\textrm{T} < 40$\;GeV/$c^2$.

The second correction, called the ``trigger-bias correction'', is required to fully account for the efficiency of the trigger selection.  The need for such correction arises from the inversion of the ratio of hadronic-to-electromagnetic energy criterion. The nonelectron $E_\textrm{T}$ multijet distribution is corrected by applying weights evaluated bin-by-bin in the control region. The weights ($w_{\rm TB}$) for each bin of the  nonelectron $E_\textrm{T}$ distribution ($E_\textrm{T}$ bin) are evaluated as follows:
\begin{equation}
\label{tbc}
w_{\rm TB}(E_\textrm{T}\ {\rm bin}) = \frac{N (E_\textrm{T} \ {\rm bin}) - n(E_\textrm{T} \ {\rm bin})}{N_{\rm MJ} (E_\textrm{T} \ {\rm bin})},
\end{equation}
where $N (E_\textrm{T} \ {\rm bin})$ is the event yield in a bin of the electron $E_\textrm{T}$ data distribution, $n(E_\textrm{T}\ {\rm bin})$ is the predicted event yield of electroweak and top-quark background events, and $N_{\rm MJ} (E_\textrm{T} \ {\rm bin})$ is the estimated number of multijet events in the same bin. To account for the possible dependence of the correction on the choice of the control region, two sets of weights from two nonoverlapping subsets of the CR, defined by the events with $m^W_\textrm{T} < 20$\;GeV/$c^2$ and by the events with $20 < m^W_\textrm{T} < 40$\;GeV/$c^2$, have been calculated. The two sets of corrections are then applied to the events populating the whole control region and the differences with respect to the nominal correction are assigned as systematic uncertainties.

After these two corrections are applied, the missing transverse energy of the event is recalculated.

In the muon channel, the multijet background is modeled using muon candidates that pass all the muon requirements~\cite{JPG342457(2007)} but with isolation~\cite{isolation} between 0.1 and 0.2, rather than  greater than 0.1, as was previously used in~\cite{JPG342457(2007)} to define nonisolated muons. In this paper these muons are referred as ``loosely-isolated muons''. Events with isolation greater than 0.2 are used to evaluate a systematic uncertainty of the model.

The multijet background yield expected in the SR ($N_{\rm MJ}|_{\rm SR}$) is estimated using the following equation: 
\begin{equation}
\label{qcdrate}
 N_{\rm{MJ}}|_{\rm SR}=  \frac{N_{\rm{MJ}}|_{\rm{CR}}}{N^*_{\rm{MJ}}|_{\rm{CR}}} \cdot N^*_{\rm{MJ}}|_{\rm{SR}},
 \end{equation}
where $N^*_{\rm{MJ}}|_{\rm{CR}}$ and $N^*_{\rm{MJ}}|_{\rm{SR}}$ are the multijet event yields in the control and in the signal regions, respectively, that pass the nonelectron or the loosely-isolated muon selections after the subtraction of contributions from processes with real leptons. The multijet yield in the CR, $N_{\rm{MJ}}|_{\rm{CR}}$, corresponds to the number of candidates in the control region ($N|_{\rm{CR}}$) minus the number of simulated ``electroweak and top-quark processes" background ($n|_{\rm{CR}}$) and signal ($N_{\rm{s}}|_{\rm{CR}}$) contributions, 

\begin{equation}
\label{h1}
N_{\rm{MJ}}|_{\rm{CR}}\hspace{-0.1cm}=\hspace{-0.1cm}(N-n-N_{\rm{s}})|_{\rm{CR}}.
\end{equation}

 \begin{figure*}[ht]	
	\begin{center}
\subfloat[]{\includegraphics[width=0.5\textwidth]{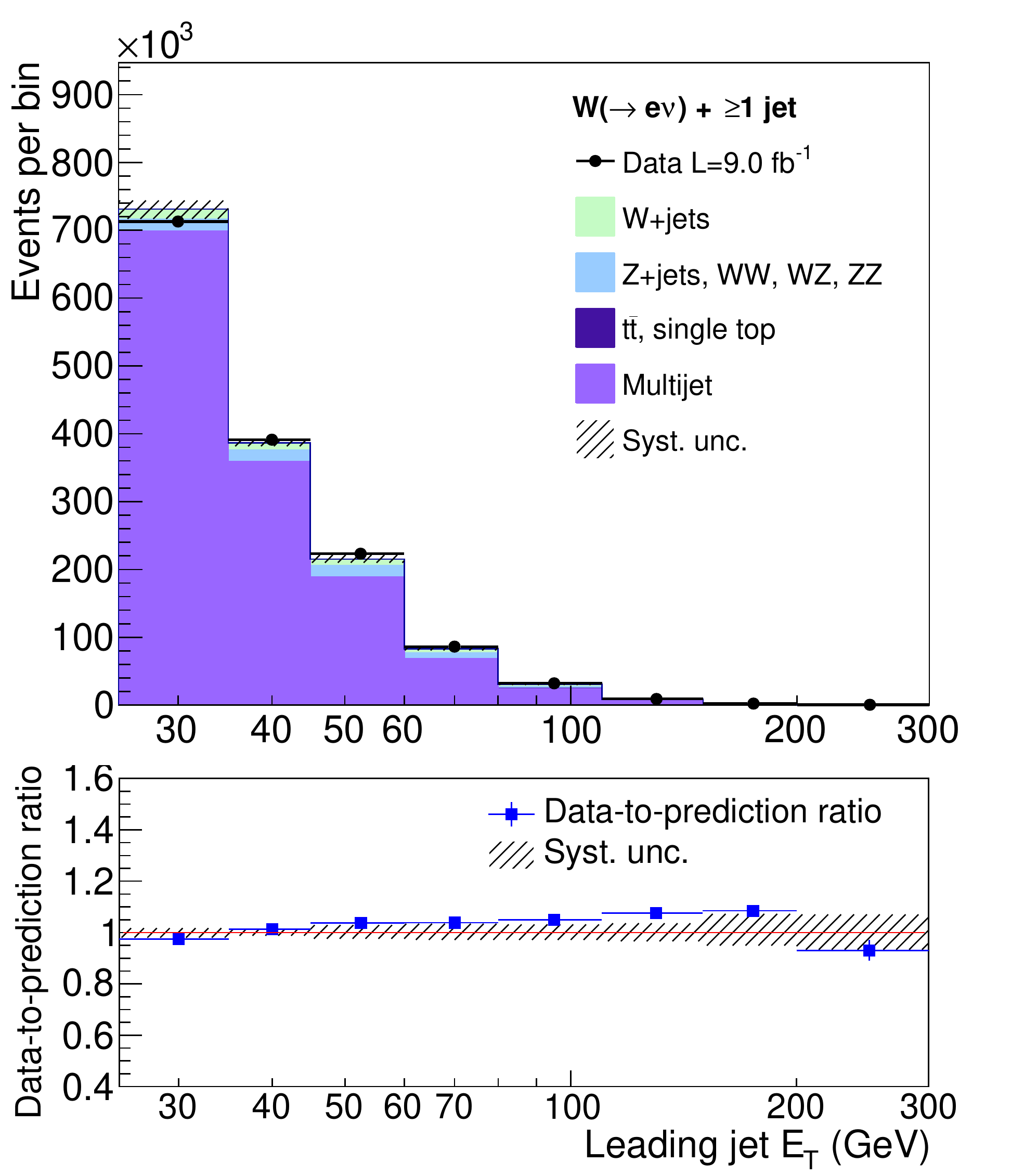}}
	\subfloat[]{\includegraphics[width=0.5\textwidth]{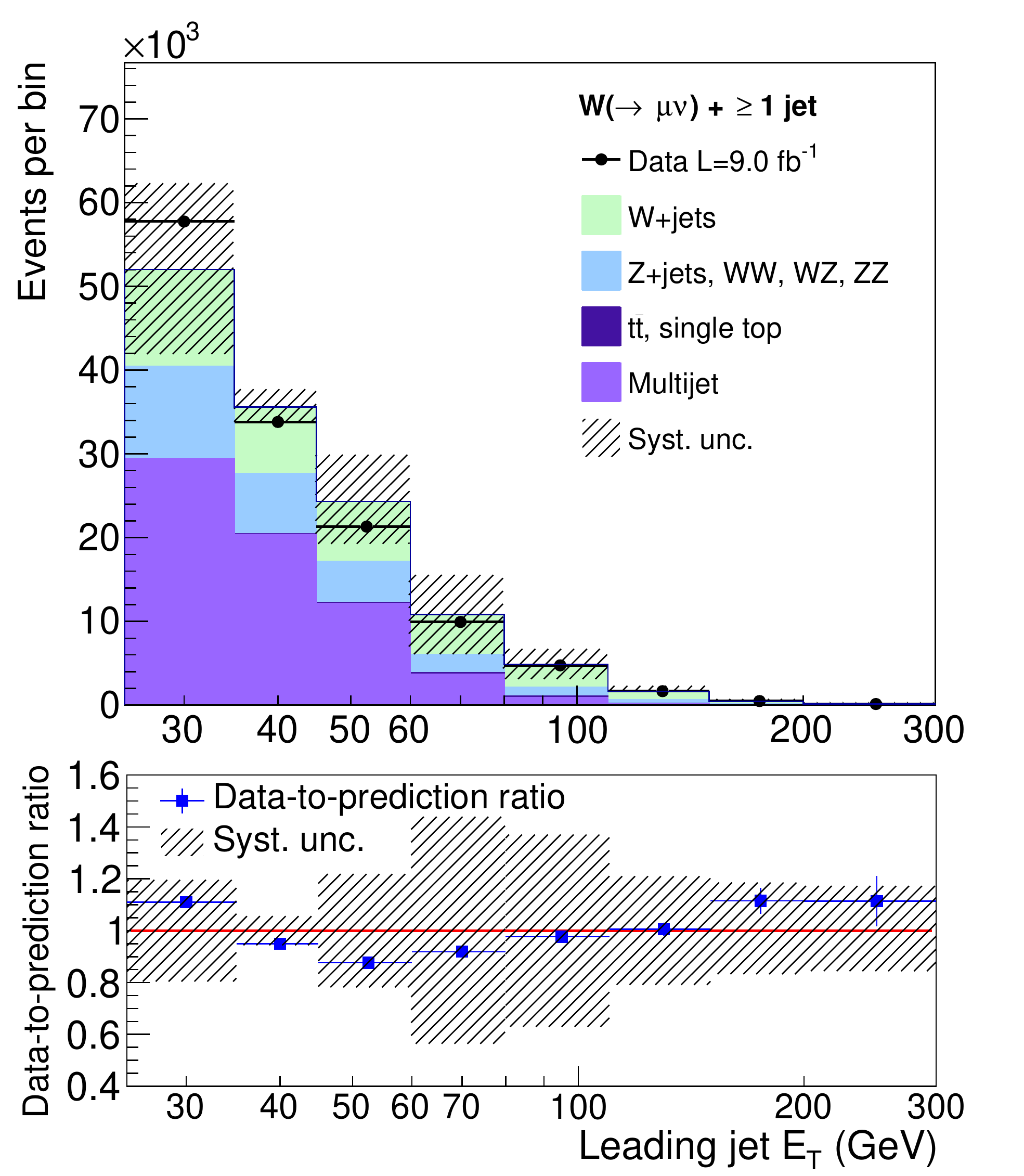}}\\	
\end{center}
 \caption{Leading jet $E_\textrm{T}$ distribution in the control region for (a) the $W(\rightarrow e \nu) + \geq 1$\,jets sample and (b) the $W(\rightarrow \mu \nu) + \geq 1$\,jets sample. 
  The data are represented with black points, while the signal and background predictions are represented with filled stacked histograms. Systematic uncertainties on the predictions are indicated by shaded areas. The lower plots show the ratios of the data to the corresponding predictions. \label{model}}
\end{figure*}

To avoid circularity, $N_{\rm{s}}|_{\rm{CR}}$ is estimated using the measured cross section $\sigma_{W+{\rm jets}}$, instead of the theoretical calculation.
The process is iterative. Starting with the approximation of a control region entirely populated of multijet events, 
\begin{equation}
\label{h0}
N_{\rm{MJ}}|_{\rm{CR}}\hspace{-0.1cm}=\hspace{-0.1cm}N|_{\rm{CR}},
\end{equation}
$\sigma_{W+{\rm jets}}$ has been calculated using the equation
\begin{equation}
\label{xsecSR}
\sigma_{W+{\rm jets}}= \frac{N|_{\rm{SR}}-n|_{\rm{SR}} - N_{\rm{MJ}}|_{\rm{SR}}}{L\mathcal{\,A}\,\epsilon},
\end{equation}
where $L  \mathcal{\,A}\, \epsilon$ is the product of the integrated luminosity, the acceptance in the SR and the total efficiency; the number $n|_{\rm{SR}}$ is the estimated yield of simulated background events in the SR, and $N_{\rm{MJ}}|_{\rm{SR}}$ is evaluated by replacing $N_{\rm{MJ}}|_{\rm{CR}}$ of Eq.\,(\ref{h0}) into Eq.\,(\ref{qcdrate}). 
The number $N_{\rm{s}}|_{\rm{CR}}$ is then calculated using Eq.\,(\ref{h1}) with $\sigma_{W+{\rm jets}}$. On the next iteration $N_{\rm{MJ}}|_{\rm{CR}}$ is then calculated with the measured value of $N_{\rm{s}}|_{\rm{CR}}$. The process is iterated until the multijet scale factor $f^{\ell}_{\rm{MJ}} =  N_{\rm{MJ}}|_{\rm{CR}}/N^*_{\rm{MJ}}|_{\rm{CR}}$ changes by less than 1\% between subsequent iterations.

 \subsection{\label{sec:IVC}Background model validation}

 The modeling of the background distributions, for both electrons and muons, is validated by comparing them with data in the CR. Examples of validation histograms are shown in Fig.~\ref{model}. Validation of the modeling of other important kinematic variables is discussed in Appendix~\ref{APPI}. The good agreement between the data and the predictions in the control region supports the validity of the background models. The shaded areas in Fig.\,\ref{model} represent the total uncertainty in the evaluation of the backgrounds previously discussed. The main systematic uncertainty in the control region is the uncertainty in the multijet model. The fractional size of this uncertainty on the control region of the muon channel is larger than that on the electron channel. The reason for this difference in uncertainties is that the identification requirements that are reversed to define the nonelectron sample have less impact on the kinematic properties of the event than the modification of the isolation requirement applied to the muon channel. 
 However, in the SR the background of the muon channel is much smaller than that in the electron channel.


 \subsection{Estimated background}
 
 \begin{table*}[t!]
  \captionsetup{labelsep=period, format=plain, justification=justified}	
  \centering 	
   \caption{Numbers of events in the data and total background for each inclusive jet multiplicity in the signal region of the electron and muon channels. 
   The individual background estimates are expressed as percentages of the numbers of events in data and are evaluated as explained in Sec.~\ref{sec:IV}.\label{tab:eve}}
   \begin{tabular*}{\textwidth}{@{\extracolsep{\fill} }lcccc}
\hline
\hline
Sample &\multicolumn{4}{c}{ $W(\rightarrow e \nu) + \geqslant N$\,jets} \\
\hline 
  Number of jets $N \geqslant$& 1 &2 &3 &4    \\
   \hline 
  {Events in data} &$477665$& $65029$& $9483$& $1642$  \\
\hline
 {Total background prediction}  &$182000\pm24000$& $30800\pm2900$& $5700\pm440$& $1320\pm110$   \\ 
Multijet &30\% &33\% &32\% &29\%  \\ 
$Z+$jets &5\% &4\% &4\% &2\% \\ 
$t\bar{t}$ &1\% &4\% &19\% &45\%  \\ 
$W(\rightarrow \tau \nu)+$jets  &2\% &1\% &1\% &1\%  \\ 
Single top-quark  &$<$1\% &2\% &3\% &2\%  \\ 
{\it WW}, {\it WZ}, {\it ZZ}  &1\% &3\% &2\% &2\%  \\ 
\hline
\hline
Sample &\multicolumn{4}{c}{ $W(\rightarrow \mu \nu) + \geqslant N$\,jets}\\
\hline 
  Number of jets $N \geqslant$& 1 &2 &3 &4    \\
   \hline 
   {Events in data} & $229823$& $28038$&$3967$ &$807$  \\
\hline
{Total background prediction}  & $39800\pm5600$& $7270\pm760$&$1860\pm150$ &$550\pm40$  \\ 
Multijet  &3\% &3\% &3\% &2\% \\
$Z+$jets  &10\% &9\% &7\% &4\% \\ 
$t\bar{t}$ &1\% &6\% &28\% &57\% \\ 
$W(\rightarrow \tau \nu)+$jets   &2\% &2\% &1\% &1\% \\ 
Single top-quark   &1\% &3\% &4\% &3\% \\ 
{\it WW}, {\it WZ}, {\it ZZ}  &1\% &4\% &4\% &2\% \\ 
\hline
\hline
\end{tabular*}
 \end{table*}

The background contributions for each inclusive jet multiplicity in the SR are summarized in Table~\ref{tab:eve}. For $N \geqslant 1$ and $N \geqslant 2$ jets, multijet production and $Z+$jets represent the main background contributions to the electron and the muon channels, respectively, while the $t\bar{t}$ background contribution is the largest single contribution for the sample with $N \geqslant 4$ jets in both channels. For $N \geqslant 3$, the main background contributions are multijet production in the electron channel and $t\bar{t}$ in the muon channel.
The contributions of the {\it WW}, {\it WZ}, {\it ZZ}, and the single-top backgrounds are largest for $N \geqslant 2-4$\,jet multiplicities but do not exceed 4\% in either channel. Table~\ref{tab:eve} reports also the number of selected events for each inclusive number of jets. 


\section{\label{sec:V}Unfolding}
 
\begin{figure*}[ht!!!!]
 \captionsetup{labelsep=period, format=plain, justification=justified}	
	\centering
	\vspace{0.0cm}	
	\subfloat[]{\includegraphics[width=0.5\textwidth]{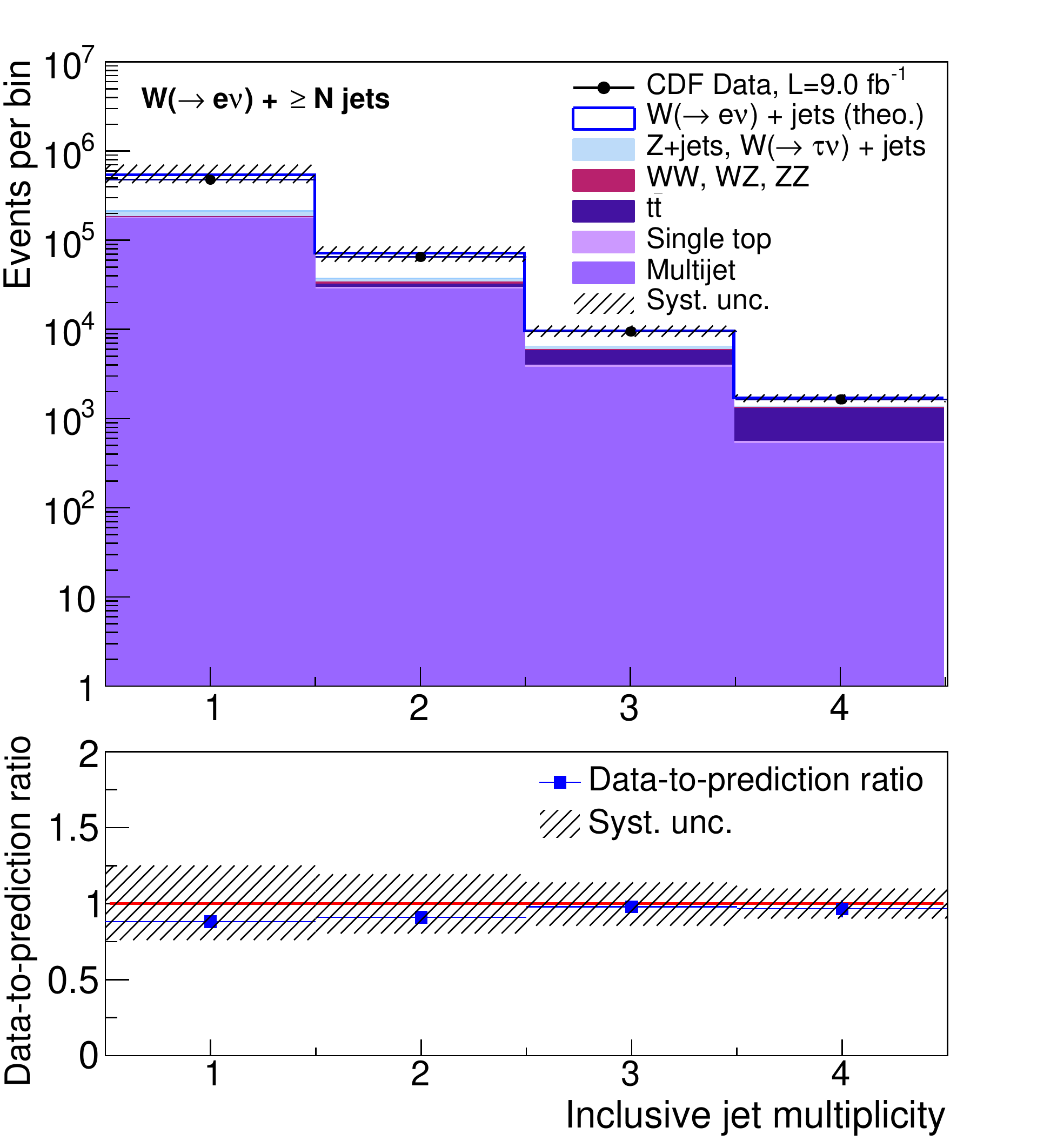}}
	\subfloat[]{\includegraphics[width=0.5\textwidth]{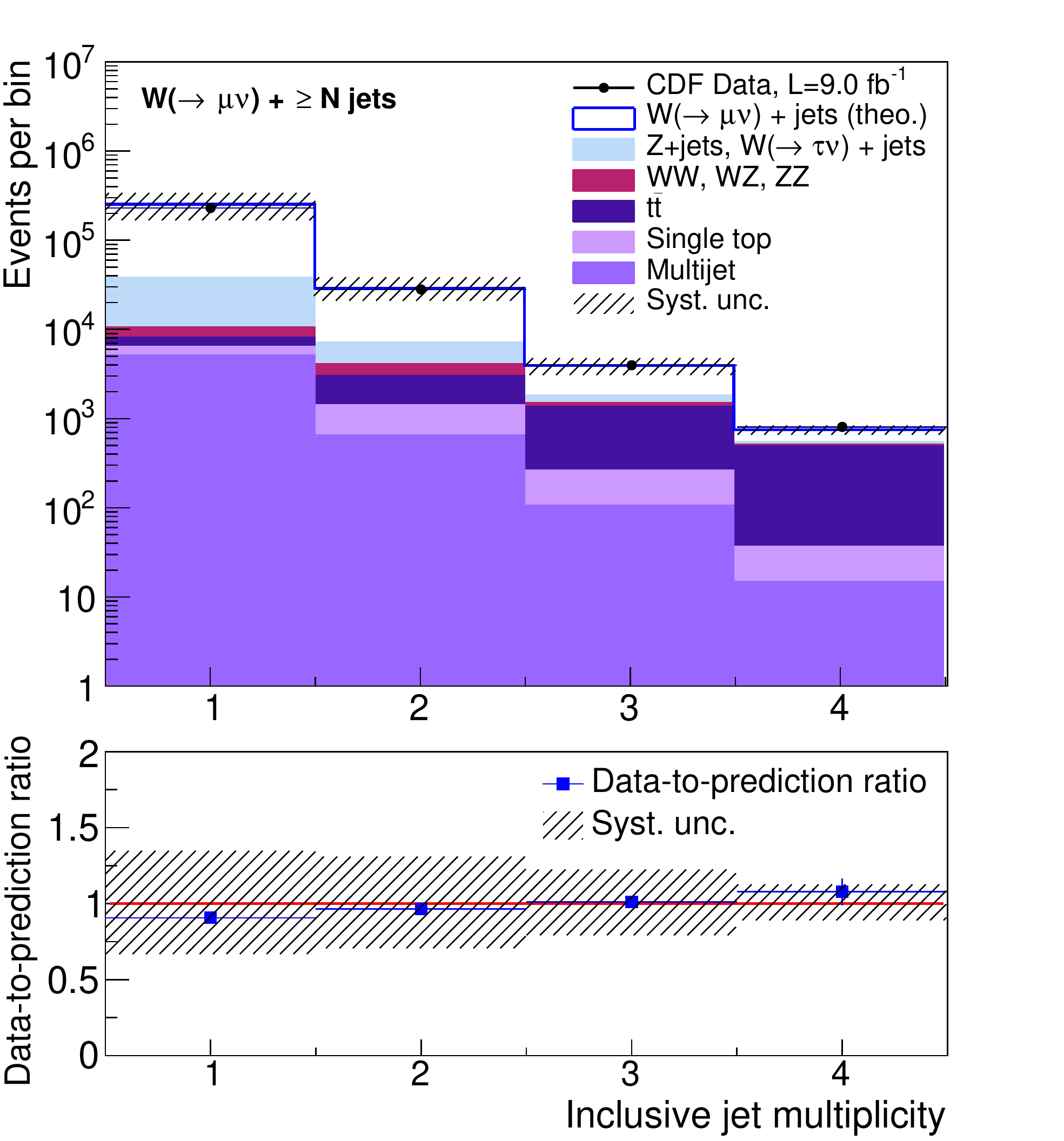}}\\	
	\vspace{0.0cm}	
	\subfloat[]{\includegraphics[width=0.5\textwidth]{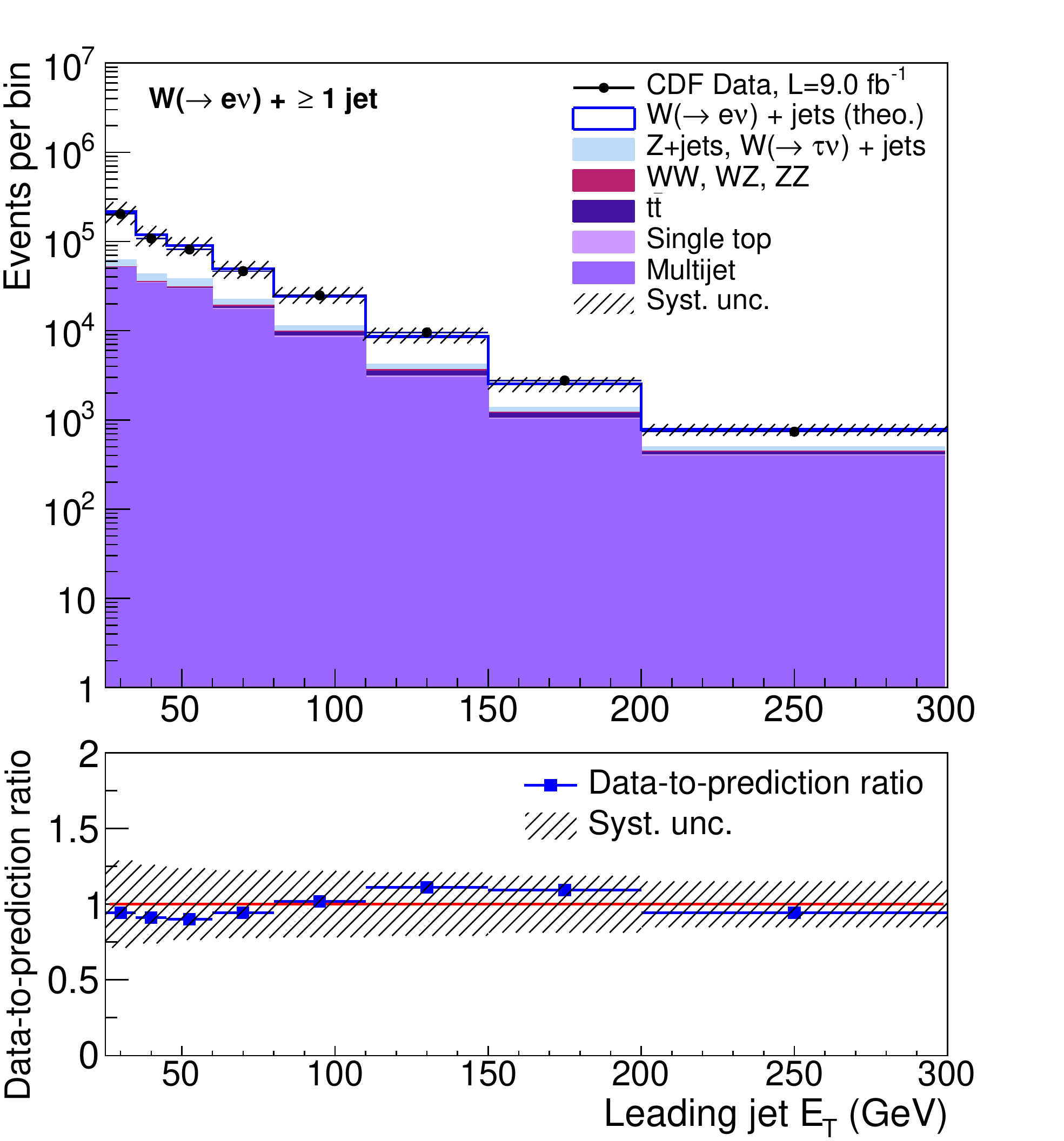}}
	\subfloat[]{\includegraphics[width=0.5\textwidth]{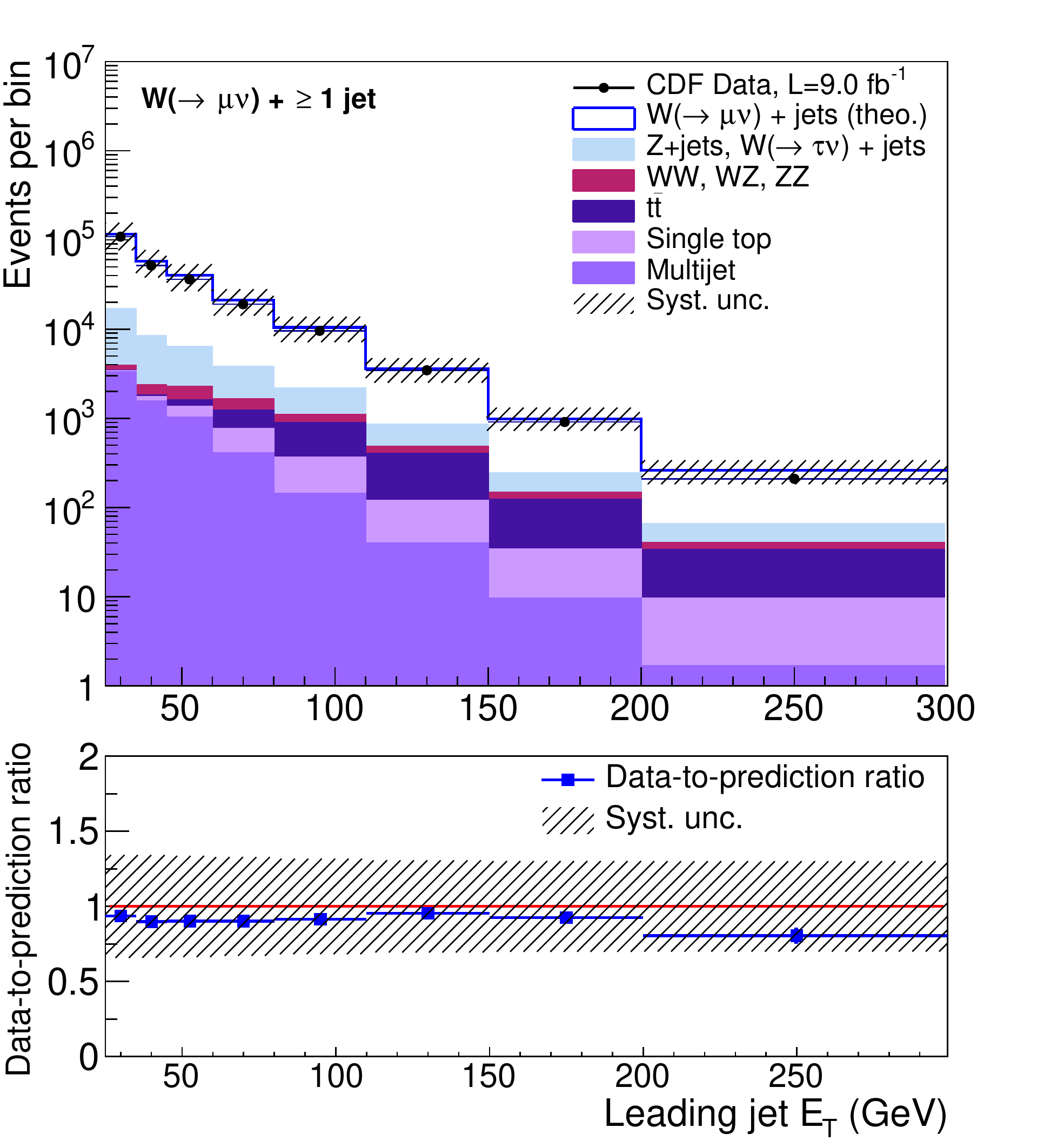}}\\	
	\vspace{0.0cm}	
	\caption{Distributions of data for the $W\!\rightarrow\! e \nu$ channel overlaid with predicted background and signal (a) for each jet multiplicity and (c) as a function of the leading-jet $E_\textrm{T}$. The same distributions are reported in (b) and (d) for the  $W\!\rightarrow\! \mu \nu$ channel. The predicted signal is obtained by using  an \textsc{Alpgen+Pythia} LO cross-section calculation multiplied by the $K$-factor. The lower plots show the ratio between data and prediction. The shaded regions represent the systematic uncertainties. \label{NSR}}
\end{figure*}

The observed and expected distributions of the inclusive jet multiplicity and the leading-jet $E_\textrm{T}$ for events passing the signal selection requirements are shown in Fig.~\ref{NSR}. The expected signal yields are predicted with a MC calculation based on \textsc{Alpgen+Pythia} propagated through the detector simulation and are normalized to the LO calculations scaled by a $K$-factor of 1.4~\cite{PRD77011108(2008)}.

The signal distributions  obtained  by subtracting the estimated background are influenced by the acceptance, nonlinear response, and finite resolution of the detector. To correct for these effects, and to facilitate comparison with theory, the distributions are unfolded back to the particle-level separately for the two channels. The particle-level leptons and jets are reconstructed from all simulated particles with a lifetime of more than 10\,ps, before the detector simulation, and by applying the same requirements as those used for the experimental data. The electron or muon from the $W$-boson decay is recombined with a radiated photon if the radial distance between the lepton and the photon ($\Delta R(\ell,\gamma$)  is less than 0.1. The neutrino momentum from the $W$-boson decay is used to calculate the missing transverse energy. Particle-level jets are constructed by applying the JETCLU algorithm with cone radius 0.4 to the stable particles, from which  the lepton (including the recombined photon) and neutrino from the $W$-boson decay are removed.
The effects of the detector are described by a response matrix, determined from simulation, that maps all the generated events (at the particle level) to the reconstructed events (at the detector level).
The response matrix for each distribution subjected to unfolding is built using the $W+$jets sample generated with \textsc{Alpgen+Pythia}, which has approximately ten times the size of the data sample.
In the case of the jet-multiplicity distribution, unfolding is performed using bins that correspond to exclusive numbers of jets, except the last bin, {\it i.e.}, the response matrix has bins corresponding  to events with a $W$ boson and exactly one, exactly two, exactly three, or at least four jets. The leading-jet $E_\textrm{T}$ distribution, for each of the two lepton channels, is unfolded considering bins designed to contain a sufficiently large number of events. The response matrices are determined by considering the jet with highest $E_\textrm{T}$ at the particle and detector levels independently, without any geometric matching between the two.

The first unfolding step consists of applying the inverted response matrix to the observed distribution. The matrix inversion is performed using the regularized singular-value decomposition (SVD) technique~\cite{NIMA372469(1996),NIMA38981(1997)}.  The SVD-inversion results are rendered robust against fluctuations of the bin populations in data and MC by introducing a regularization condition, namely a ``minimum curvature condition''~\cite{NIMA372469(1996)} on the unfolded distribution, to avoid amplifying fluctuations coming from sparsely-populated MC and data bins.
Regularization is characterized by a strength parameter. In this analysis, each distribution is unfolded by optimizing its strength parameter, as explained in Ref.~\cite{NIMA372469(1996)}.  In addition, it has been verified  that the method employed in choosing each regularization parameter leads to a value that introduces the lowest systematic bias into the unfolded distribution. This is done using test distributions similar to the observed distributions. The systematic uncertainty on the residual bias in the unfolding procedure is discussed in the next section.

The unfolded event phase space has been restricted to the fiducial region by applying an acceptance matrix after applying the inverse of the response matrix. The acceptance matrix is determined using the $W+$jets sample generated with \textsc{Alpgen+Pythia} for each unfolded distribution. The response and acceptance matrices are reported Appendix~\ref{APPII}.

\section{\label{sec:VI} Systematic Uncertainties}

\begin{figure*}[ht!]
    \captionsetup{labelsep=period}	
	\centering
	\subfloat[]{\includegraphics[width=0.5\textwidth]{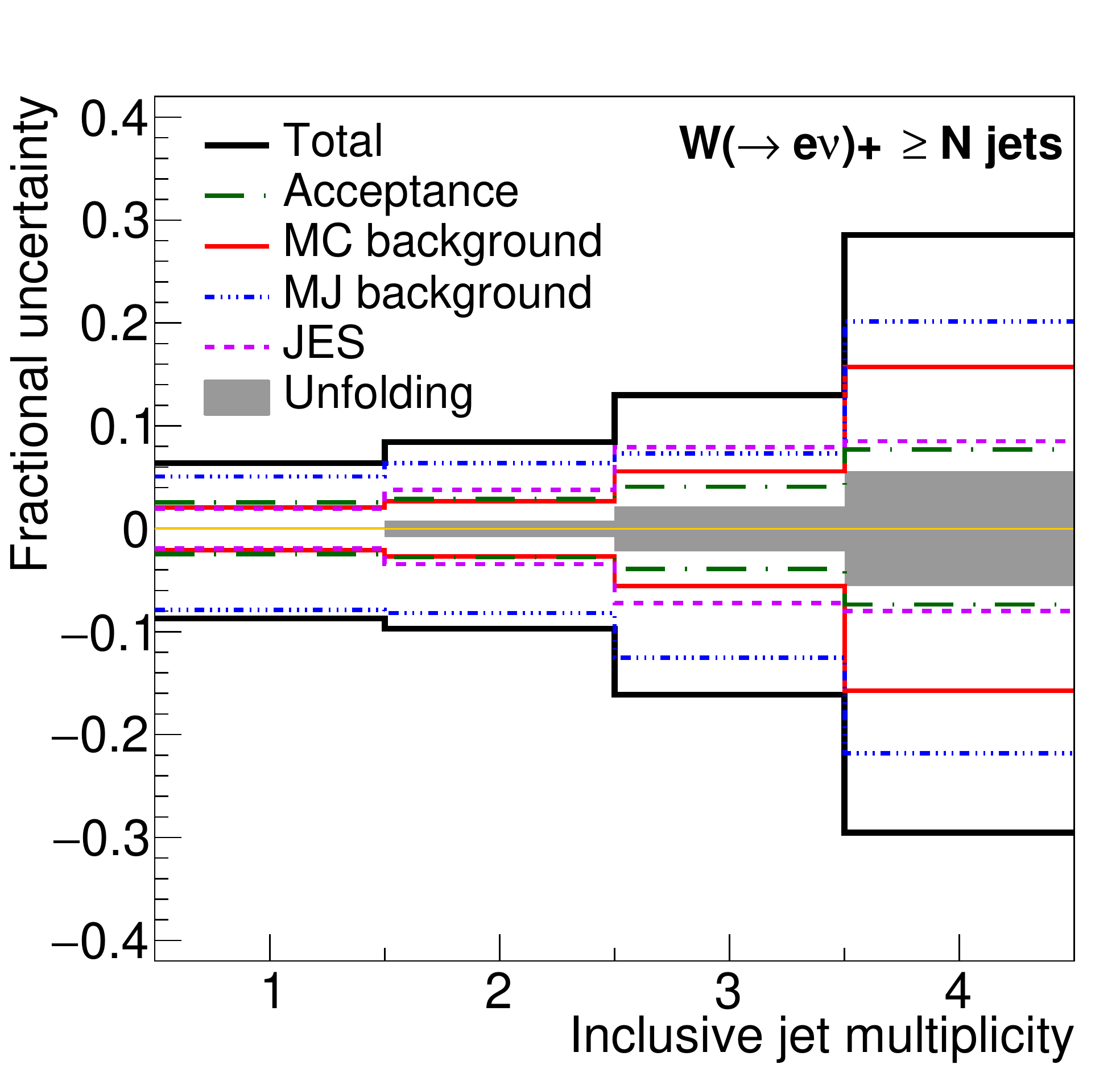}}
	\subfloat[]{\includegraphics[width=0.5\textwidth]{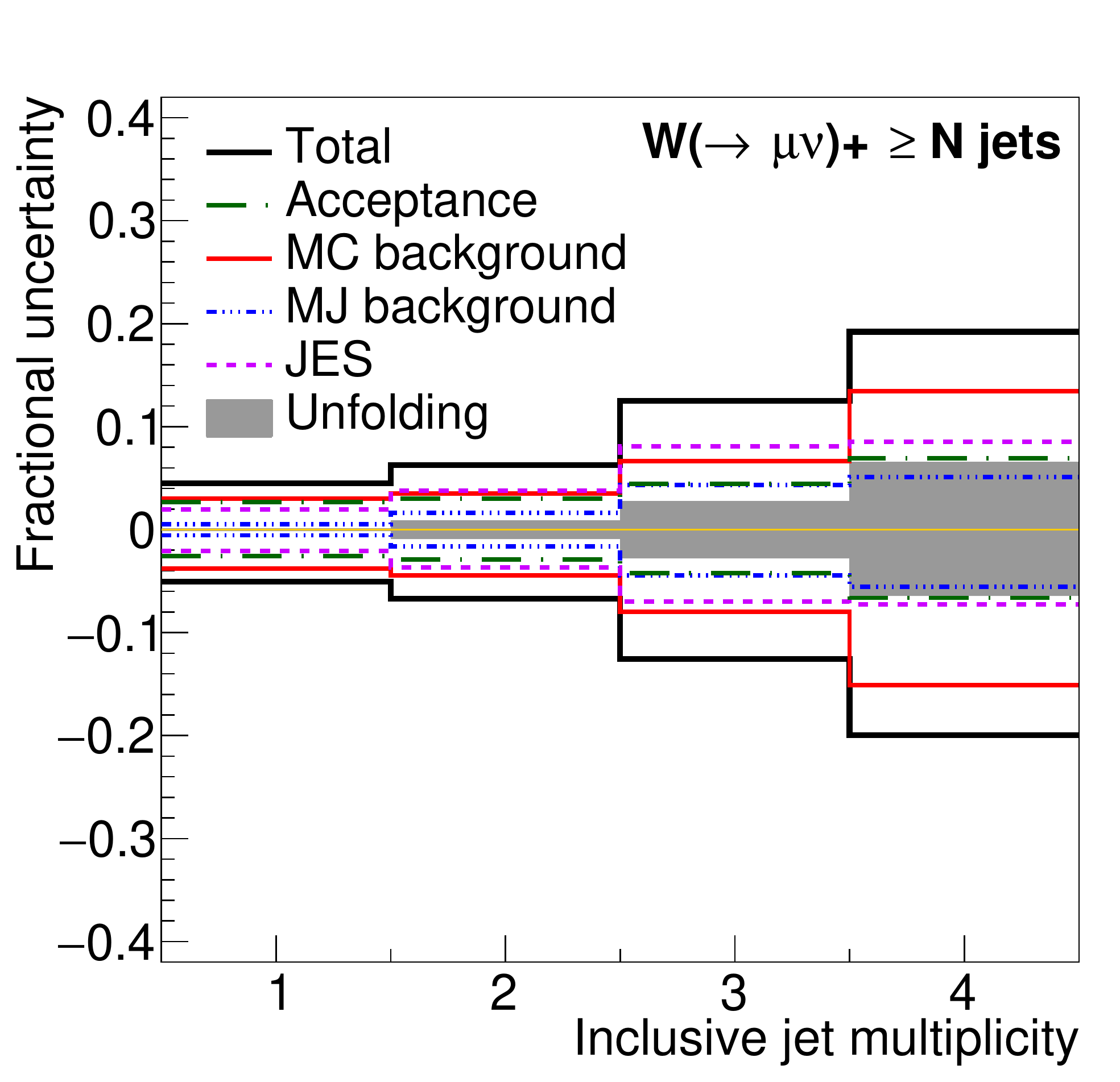}}\\	
	\caption{Fractional systematic uncertainties as functions of inclusive jet multiplicity (a) in the $W\!\rightarrow\! e \nu$ channel and (b) in the $W\!\rightarrow\! \mu \nu$ channel.\label{SYS_nj}}
	\vspace{0.0cm}
	\centering
	\subfloat[]{\includegraphics[width=0.5\textwidth]{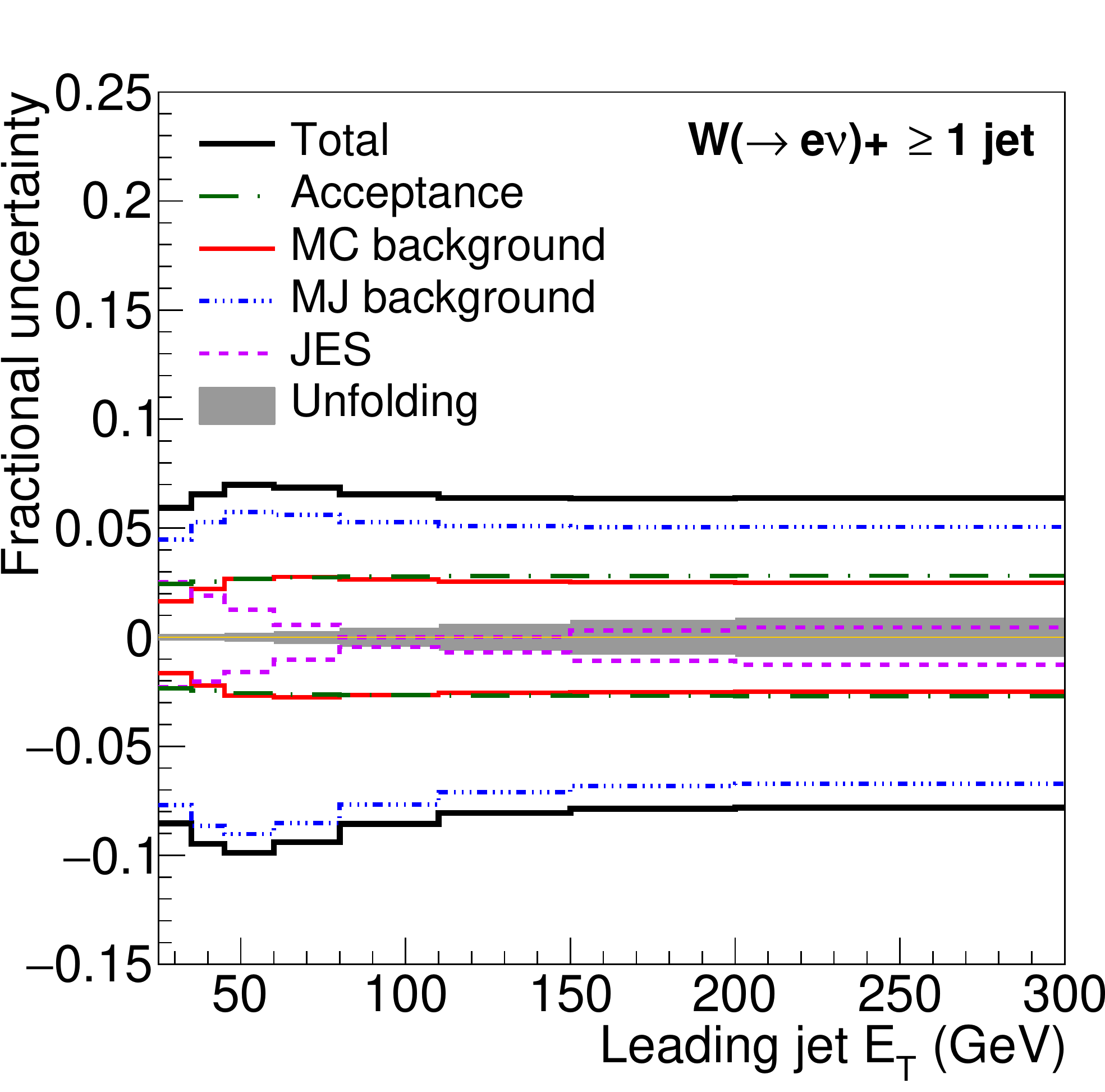}}
	\subfloat[]{\includegraphics[width=0.5\textwidth]{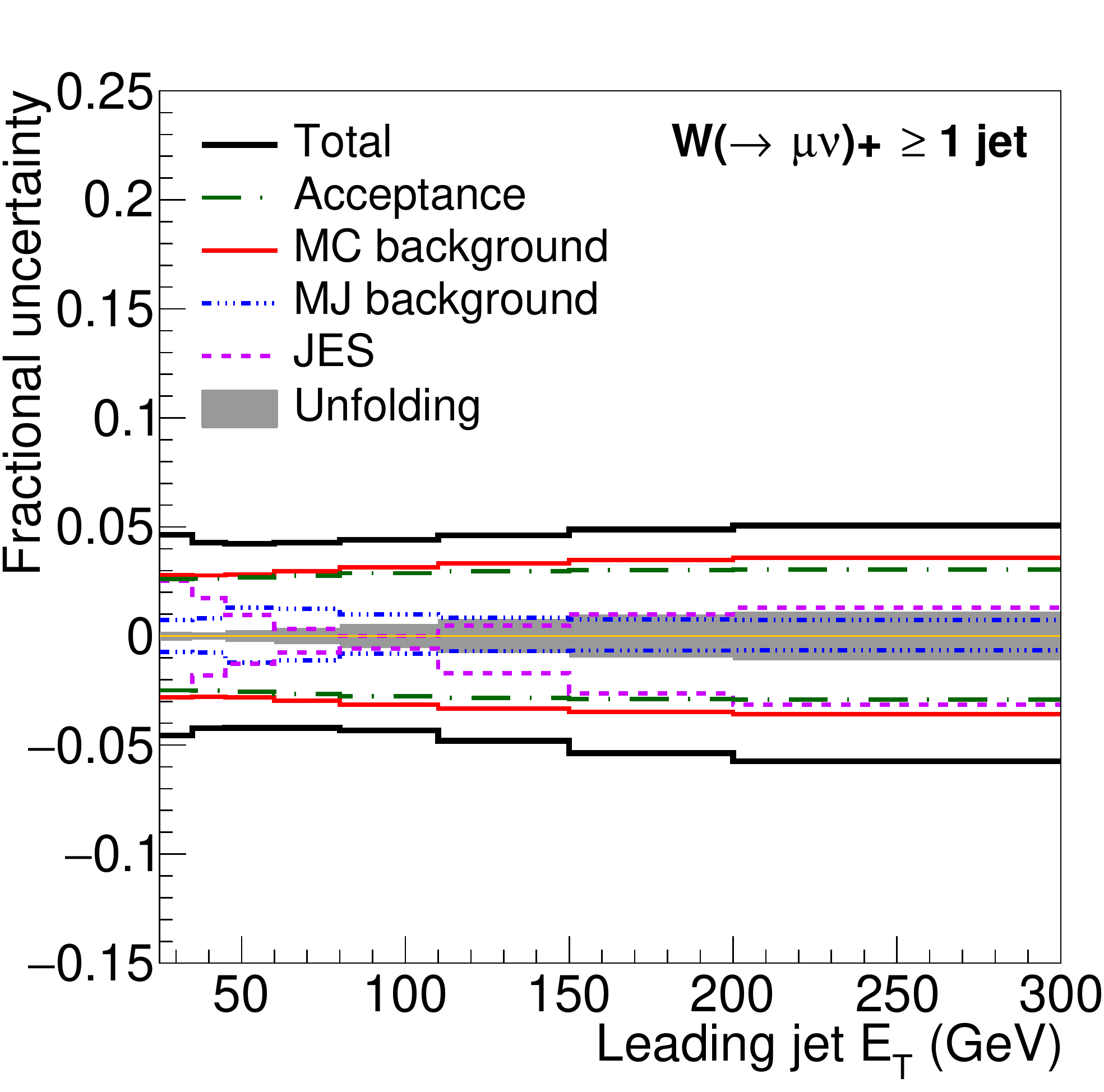}}\\	
\caption{Fractional systematic uncertainties as functions of leading jet $E_\textrm{T}$ (a) in the $W\!\rightarrow\! e \nu$ channel and (b) in the $W\!\rightarrow\! \mu \nu$ channel.\label{SYS_j1et}}
\end{figure*}

The systematic uncertainty on each of the unfolded measurements is assessed by repeating the unfolding procedure for variations of each parameter associated with systematic effects. Figures \ref{SYS_nj} and \ref{SYS_j1et} show the systematic uncertainties obtained as differences between each set of unfolded data and the nominal result. 

These uncertainties include the contributions, discussed in Sec.~\ref{sec:IV}, from the simulated background normalizations, the lepton acceptances, the jet-energy scale (JES), and the estimated multijet background yield.

In the electron channel, the dominant source of systematic uncertainty arises from the multijet background estimate. As discussed in Sec.~\ref{sec:IV}, the multijet background is determined from data and validated using the CR. The uncertainty on this model includes contributions from the corrections applied to the data-driven sample as well as contributions from the multijet scale factor $f^{e}_{\rm MJ}$. The corrections applied to the model are the nonelectron energy-scale correction and the trigger-bias correction. The uncertainties of the nonelectron energy-scale correction are obtained by using the correction factors shifted by one standard deviation (10\%--13\%) and by smearing the missing transverse energy of the event (2\%--13\%), while those of the trigger-bias correction are derived by comparing data and predictions in two control regions restricted to $m^W_\textrm{T} < 20$\;GeV/$c^2$ and $20 < m^W_\textrm{T} < 40$\;GeV/$c^2$ (1\%--5\%). The uncertainty on $f^{e}_{\rm MJ}$ is due to the contributions from CR background suppression (2\%--5\%). The overall uncertainty on the inclusive cross sections ranges from 6\% for $W(\rightarrow e\nu)+ \geqslant 1$\,jet results to 21\% for $W(\rightarrow e\nu)+ \geqslant 4$\,jets results.

In the muon channel, the major source of systematic uncertainty is the uncertainty on the MC prediction of the background (3\%--14\% of the $W(\rightarrow \mu\nu)+\geqslant N$\,jets cross sections for $N = 1-4$). This includes the uncertainty on the theoretical cross section used to normalize each background process (6\% for the {\it WW}, {\it WZ}, and {\it ZZ} processes, 3\% for $t\bar{t}$ production, 11\% for single-top-quark production, 20\% and 40\% for $Z$+jets and $W(\rightarrow\tau\nu$+jets) processes, respectively), and 6\% uncertainty on the integrated luminosity of the sample. In this channel, the uncertainty on the multijet background estimate gives the lowest contribution (1\%--5\%) to the overall uncertainty on the cross sections and is determined by varying the isolation requirement.

Additional sources of uncertainty are the contributions from electron and muon acceptances (2.2\% in both channels) and of the JES.  The impact of the JES uncertainty is estimated by computing the cross section with the JES factors shifted by one standard deviation. The difference in the scale factors for simulated gluon and quark jets is also included~\cite{NIMA566375(2006),QG}.
The resulting uncertainties on the inclusive cross sections range from about 1\% for $W+\geqslant 1$\,jet events to 11\% for $W+\geqslant 4$\,jets events in both the electron and the muon channels.

The uncertainties in the inclusive cross sections from the unfolding procedure range from 0.1\% to 7\% for inclusive jet multiplicities between one and four, respectively.
This uncertainty consists of a component due to approximations associated with the unfolding method and a component resulting from potential mismodelings in simulated $W+$\,jets events used to determine the unfolding matrix. 
The residual bias in the unfolding procedure, described in in Sec.\,\ref{sec:V}, is evaluated for each differential cross section value using simulated experiments. For each observed distribution, 1000 test distributions are generated and unfolded. The resulting distributions of the differences between generated and unfolded values are assumed to be a measure of the bias and are used as an estimate of the uncertainties of the unfolding procedure. The uncertainty in the simulated $W+$jets sample comprises the uncertainties in the lepton acceptance (2.2\%) and the uncertainties in the JES.

The total uncertainty on the unfolded inclusive cross sections ranges from 7\% for $W(\rightarrow e\nu)+ \geqslant 1$\,jet events to 34\% for $W(\rightarrow e\nu)+ \geqslant 4$\,jets events and from 5\% for $W(\rightarrow \mu\nu)+ \geqslant 1$\,jet results to 24\% for $W(\rightarrow \mu\nu)+ \geqslant 4$\,jets results. Table~\ref{SYST} reports a summary of the systematic uncertainties.

\begin{table*}[ht!]
 \captionsetup{labelsep=period, format=plain, justification=justified}	
  \centering 
  \caption{Summary of the systematic uncertainties. The uncertainties are listed as ranges where the impact of the uncertainty depends on the jet multiplicity. If the uncertainty has an impact on the shape of the leading jet $E_\textrm{T}$ distribution, a checkmark symbol is placed in the column labeled ``shape''. ``EW and top-quark processes'' refers to all the processes simulated with MC techniques, $Z$+jets, $W(\rightarrow \tau \nu)$+jets, {\it WW}, {\it WZ}, {\it ZZ}, $t\bar{t}$, and single top-quark.\label{SYST}}
    \begin{tabular*}{\textwidth}{@{\extracolsep{\fill} }lccc}
\hline
\hline
Source & Rate & Shape & Process affected\\
\hline
 Lepton acceptance & 2.2\% & -- & EW and top-quark processes \\
  MC background & &  &  \\
  \hspace{0.4cm}$Z$+jets normalization & 20\% & --  & $Z$+jets \\
  \hspace{0.4cm}$W(\rightarrow \tau \nu)$+jets normalization & 40\% & --  &  $W(\rightarrow \tau \nu)$+jets \\
 \hspace{0.4cm}$t\bar{t}$ normalization  & 3\% &  -- & $t\bar{t}$ \\
 \hspace{0.4cm}Single top-quark normalization  & 11\%&  -- &  Single top-quark \\
 \hspace{0.4cm}{\it WW}, {\it WZ} and {\it ZZ}  normalization & 6\% & -- & {\it WW}, {\it WZ}, {\it ZZ} \\
 \hspace{0.4cm}Luminosity & 6\% & --& EW and top-quark processes$$\\
  MJ background & &  &  \\
  \hspace{0.4cm}Statistical uncertainty & $0.1\%-8$\% ($1\%-37$\%)& --  & Multijet electron (muon) sample \\
  \hspace{0.4cm}Multijet Scale factor ($f^{\ell}_{\rm{MJ}}$)  &$2\%-5$\% ($13\%-68$\%)&  -- &Multijet electron (muon) sample  \\
\hspace{0.4cm}Nonelectron energy scale & $10\%-13$\% & \checkmark & Multijet electron sample \\
\hspace{0.4cm}Nonelectron energy resolution & $1\%-13$\% & \checkmark & Multijet electron sample \\
 \hspace{0.4cm}Trigger bias correction  & $1\%-5$\% & \checkmark & Multijet electron sample   \\
 \hspace{0.4cm}Isolation  requirement   & $1\%-5$\% & \checkmark & Multijet muon sample   \\
  Jet-energy scale & $1\%-11$\% & \checkmark  &  All backgrounds \\
  Quark/gluon JES   & $\pm2.7\%/\mp4.4$\%  & -- &  EW and top-quark processes \\
  Unfolding  & $0.1\%-7$\% & \checkmark  & $W+$jets  \\
  Luminosity & 6\% & -- & $W+$jets cross section \\
    \hline
\hline
\end{tabular*}
\end{table*}

\section{\label{sec:VII} Channel Combination}

\begin{figure*}[th!]
 \captionsetup{labelsep=period, format=plain, justification=justified}	
	\centering
	\subfloat[]{\includegraphics[width=0.5\textwidth]{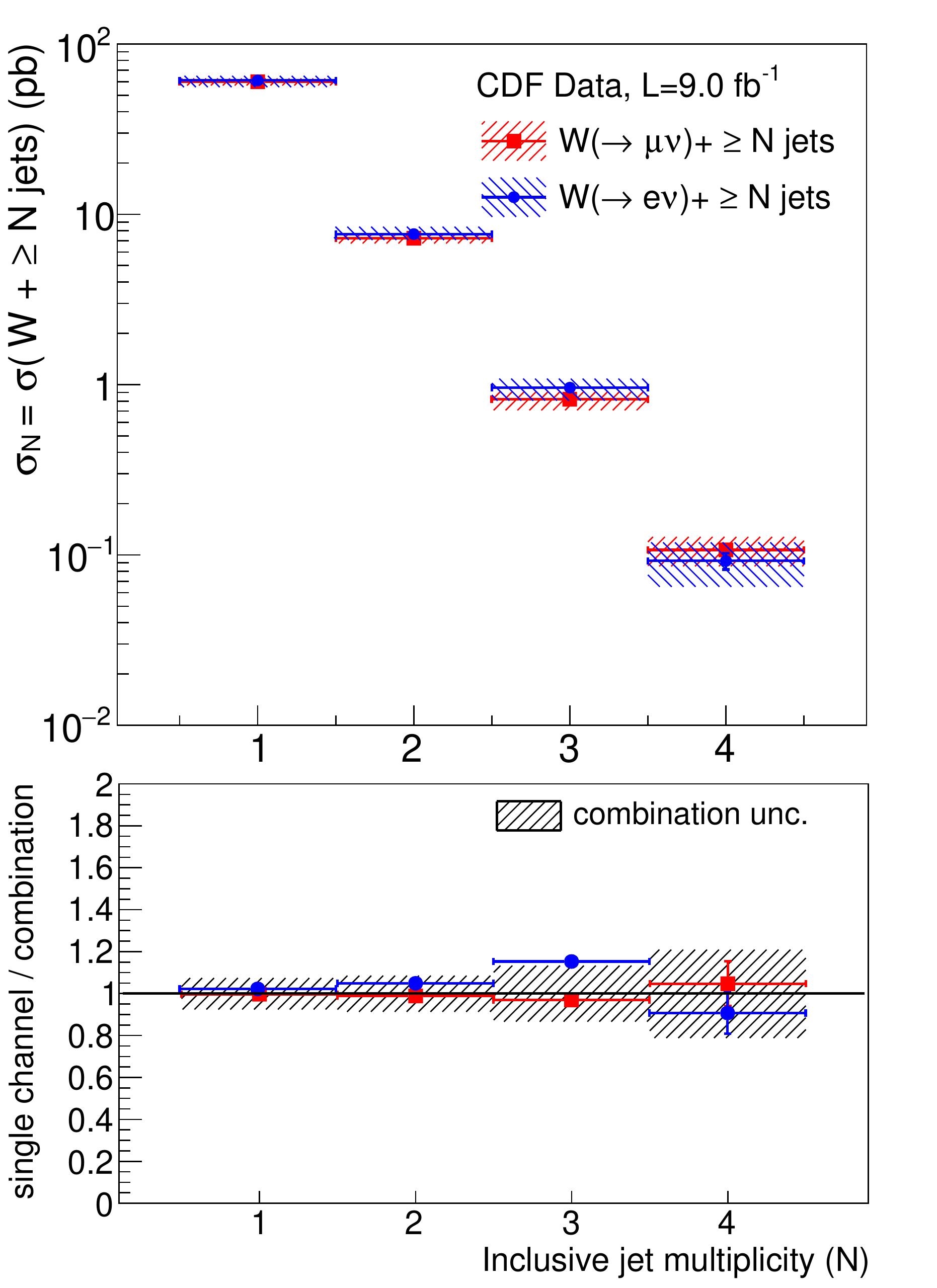}}
	\subfloat[]{\includegraphics[width=0.5\textwidth]{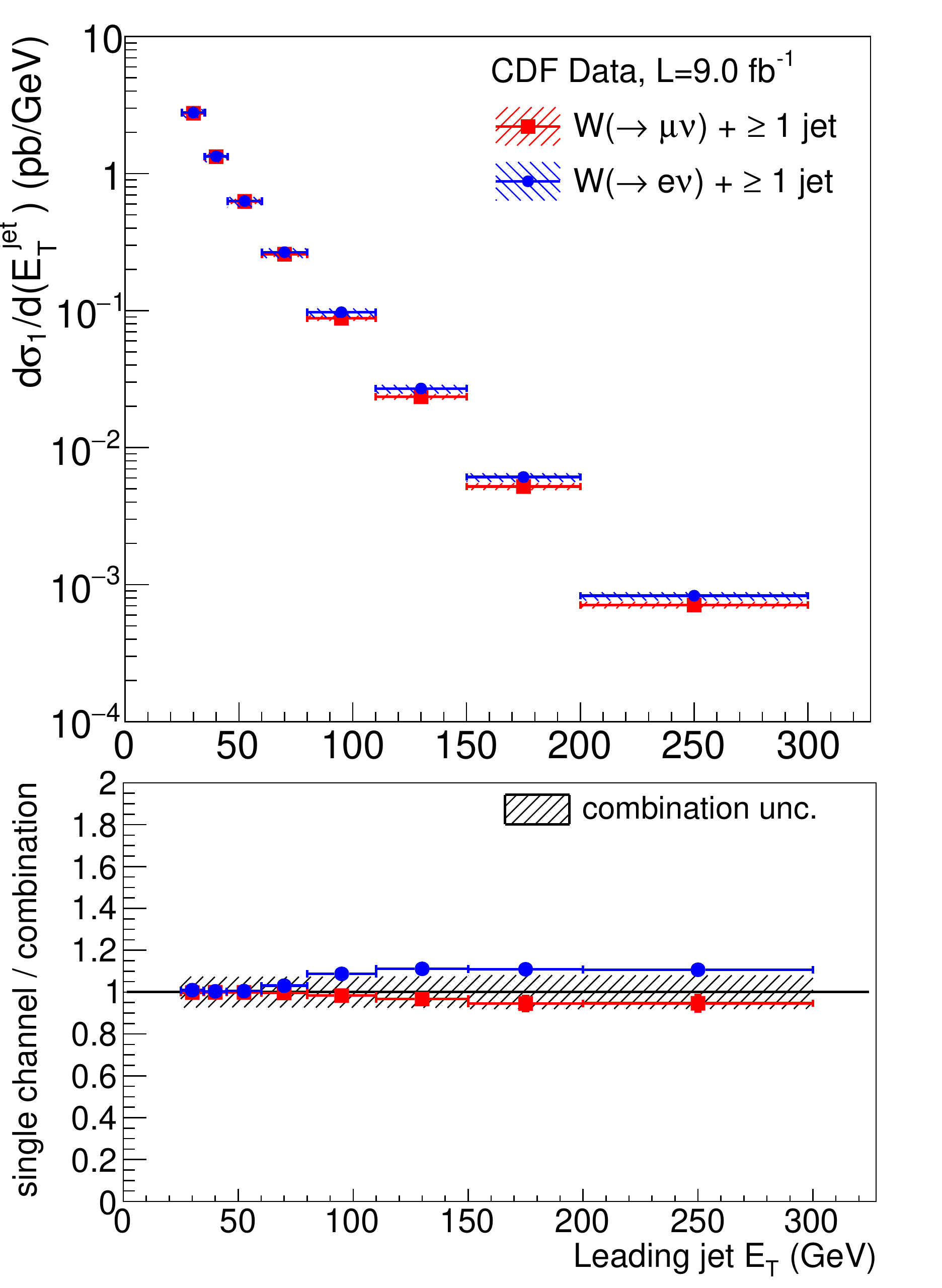}}\\	
	\caption{Measured cross sections in the $W(\rightarrow e \nu)+$\,jets (dots) and $W(\rightarrow \mu \nu)+$\,jets (squares) decay channels as functions of (a) the inclusive number of jets and (b) the inclusive leading jet $E_\textrm{T}$ in events with one or more jets. The hatched areas in the upper plots are the total uncertainties, reported  separately for the electron and muon channels. 
	The lower plots show the ratios between measured cross sections and the combined results. The bands correspond to the systematic uncertainties of the combination. \label{AIB}}
\end{figure*}

The cross sections are calculated by dividing the signal yields resulting from the unfolding by the integrated luminosity. The measurements from the muon and electron channels are combined. Assuming lepton universality, the combination is performed by using the asymmetric iterative best linear unbiased estimate method \cite{NIMA270110(1988),ICHEP08}. This method accounts for the correlations of asymmetric uncertainties. Systematic uncertainties related to JES, MC-based predictions, unfolding and luminosity are considered  to be 100\% correlated between channels. Statistical, acceptance, and multijet background uncertainties are considered uncorrelated between muons and electrons. The upper plots in Fig.\,\ref{AIB} show the observed inclusive and differential cross sections multiplied by the branching fractions of $W$-boson decays into electrons or muons. The ratios and the combined results are shown in the lower panels. The uncertainties on the ratios are small because of correlations between the uncertainties on the individual and combined results. A detailed test of consistency with lepton universality follows.

Assuming that the couplings of the $W$ bosons in the electroweak and top-quark background processes are those predicted by the standard model, the ratio $|g_{\mu}|/|g_{e}|$ is given by the ratio of the cross sections measured in the two channels,

\begin{equation}
\label{universality}
\frac{\sigma_{W+{\rm jets}}  \ {\cal B}(W\rightarrow \mu \nu)}{\sigma_{W+{\rm jets}} \ {\cal B}(W\rightarrow e \nu)} =  \frac{\Gamma(W\rightarrow \mu \nu)}{\Gamma(W\rightarrow e \nu)} = \frac{g^2_{\mu}}{g^2_{e}},
 \end{equation}
where $\sigma_{W+{\rm jets}}$ is the inclusive production cross section, ${\cal B}(W\rightarrow \mu \nu)$ and ${\cal B}(W\rightarrow e \nu)$ are the branching fractions and $\Gamma(W\rightarrow \mu \nu)$ and $\Gamma(W\rightarrow e \nu)$ are the decay widths of the $W$ boson in the muon and the electron channels, respectively.

The resulting values of $|g_{\mu}|/|g_{e}|$ for each jet multiplicity are reported in Table~\ref{U}. The magnitudes of coupling ratios for various jet multiplicities are consistent with the previous CDF measurement, $0.991 \pm 0.012$, obtained in the inclusive channel, {\it i.e.}, $W+\geqslant 0$\,jets~\cite{JPG342457(2007)}. Consistency between results of inclusive jet multiplicities measured here with those measured previously and with lepton universality promote confidence in the cross section measurements and support channel combination.

\begin{table}[h!]
 \captionsetup{labelsep=period, format=plain, justification=justified}	
  \centering
   \caption{Magnitude of the ratio of $W\rightarrow \ell \nu$ coupling constants, $|g_{\mu}|/|g_{e}|$, measured from the ratio of the  $W(\rightarrow \mu \nu)+$jets  and $W(\rightarrow e \nu)+$jets cross sections for each inclusive jet multiplicity.\label{U}}
   \begin{tabular*}{\columnwidth}{@{\extracolsep{\fill} }lc}
\hline
\hline
Jet multiplicity & $|g_{\mu}|/|g_{e}|$ \\
 \hline
\multirow{2}{*}{$\geqslant 1$\,jet} & \multirow{2}{*}{{0.992 $\pm$ 0.002 (stat)} $^{+0.050}_{-0.053}$ (syst)}\\
& \\
\multirow{2}{*}{$\geqslant 2$\,jets} & \multirow{2}{*}{0.972 $\pm$ 0.006 (stat)  $^{+0.060}_{-0.064}$ (syst)}\\
& \\
\multirow{2}{*}{$\geqslant 3$\,jets} & \multirow{2}{*}{0.918 $\pm$ 0.020 (stat)  $^{+0.099}_{-0.116}$ (syst)}\\
& \\
\multirow{2}{*}{$\geqslant 4$\,jets} & \multirow{2}{*}{1.077 $\pm$ 0.076 (stat)  $^{+0.203}_{-0.243}$ (syst)}\\
& \\
 \hline
  \hline
\end{tabular*} 
 \end{table}

\section{\label{sec:VIII} Theoretical Predictions}

The differential cross section measurements are compared with the predictions from various theoretical calculations. 

The \textsc{Alpgen+Pythia} predictions use \textsc{Alpgen}~\cite{alpgen} to simulate the production of $n$ partons in association with a $W$ boson and \textsc{Pythia}~\cite{pythia} to perform the showering and hadronization. The CTEQ5L~\cite{Lai:1999wy} parton distribution functions are used with the nominal choice of the renormalization and factorization scale ($\mu$), {\it i.e.,} $\mu_0=\sqrt{m^2_W+P_\textrm{T}^2}$, where $P_\textrm{T}^2$ is the sum of the squared transverse momenta of all final-state partons from the same interaction point.
Jets are clustered with the JETCLU algorithm with a radius parameter of 0.4.

Predictions computed with the \textsc{Mcfm 6.8}~\cite{MCFM} generator are carried out at NLO for inclusive cross sections with number of jets $N=1$ and $2$ but are limited to LO for  $N=3$. No prediction is available for $N=4$. The \textsc{Mcfm} predictions are generated using the CTEQ6.6 PDF set~\cite{cteq6.6}, with the same choice of renormalization and factorization scales as the \textsc{Alpgen+Pythia} predictions, with the exception of the LO prediction, for which $\mu_0$ equals $H_\textrm{T}/2$, where $H_\textrm{T}$ is the scalar sum of the transverse momenta of all final-state particles.

\begin{figure*}[t!]	
 \captionsetup{labelsep=period, format=plain, justification=justified}	
 \begin{center}	
{\includegraphics[width=0.95\textwidth]{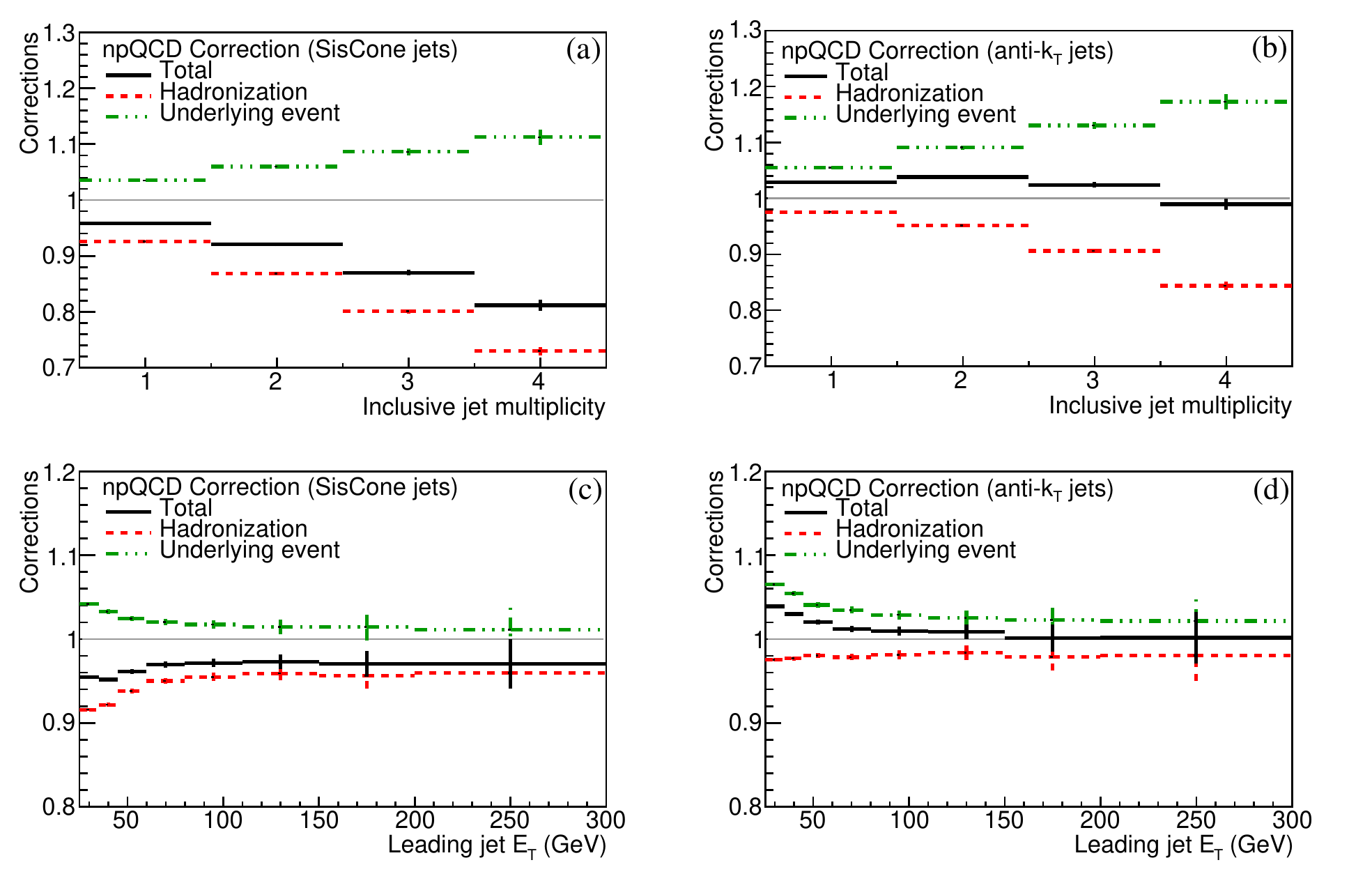}}\
	\end{center}	
	\caption{Nonperturbative QCD (npQCD) corrections applied to the \textsc{Mcfm} theoretical predictions. The corrections as functions of the inclusive jet multiplicity ((a) and (b)) and the leading jet $E_\textrm{T}$ ((c) and (d)) are derived using SisCone jets ((a) and (c)) and anti-$k_\textrm{T}$ jets ((b) and (d)). The hadronization (dashed lines) and the underlying event (dot-dashed lines) contributions to the total corrections (solid lines) are shown.\label{npQCD}}
	\vspace{0.4cm}
\end{figure*}

For comparison of the \textsc{Mcfm} predictions with measured cross sections at the particle level, hadronization is introduced using \textsc{Alpgen+Pythia} and jets are clustered at both particle level and parton level using infrared- and collinear-safe algorithms~\cite{IRCS}.  Nonperturbative QCD (npQCD) corrections due to hadronization and the underlying event~\cite{UE} are included in the \textsc{Mcfm} predictions before comparison with the measured cross sections. These npQCD corrections are shown in Fig.\,\ref{npQCD}, where the contribution due to hadronization is opposite to that of the underlying event.
The hadronization component is estimated by comparing the cross sections at the particle level with the ones at the parton level. The effects of the underlying event are evaluated by comparing cross sections at the particle level with and without the underlying event contribution. Jets are clustered at the parton level and at the particle level using both the anti-$k_\textrm{T}$~\cite{PLB641(2006)} and the SisCone v.1.4.0-devel \cite{JHEP05(2007)086} infrared- and collinear-safe algorithms provided by the {\tt FastJet} v.2.4.1 package~\cite{FastJet}. In both algorithms the jet radius is set to be 0.4 and for the SisCone algorithm two jets are merged when they share 75\% of energy.

The systematic uncertainties in the theoretical cross sections contain contributions from PDF and scale uncertainties. The PDF uncertainties are obtained by varying each of the eigenvalues in the CTEQ set by plus or minus one standard deviation~\cite{HESS}. The largest uncertainty in the theoretical predictions is due to the choice of the renormalization and factorization scales, which are kept equal to each other and are varied between the extremes $\mu_0/2$ and $2\mu_0$.

\section{\label{sec:IX} Results and Comparison with theoretical predictions}

\begin{figure*}[ht!]	
 \captionsetup{labelsep=period, format=plain, justification=justified}	
 \begin{center}	
{\includegraphics[width=\textwidth]{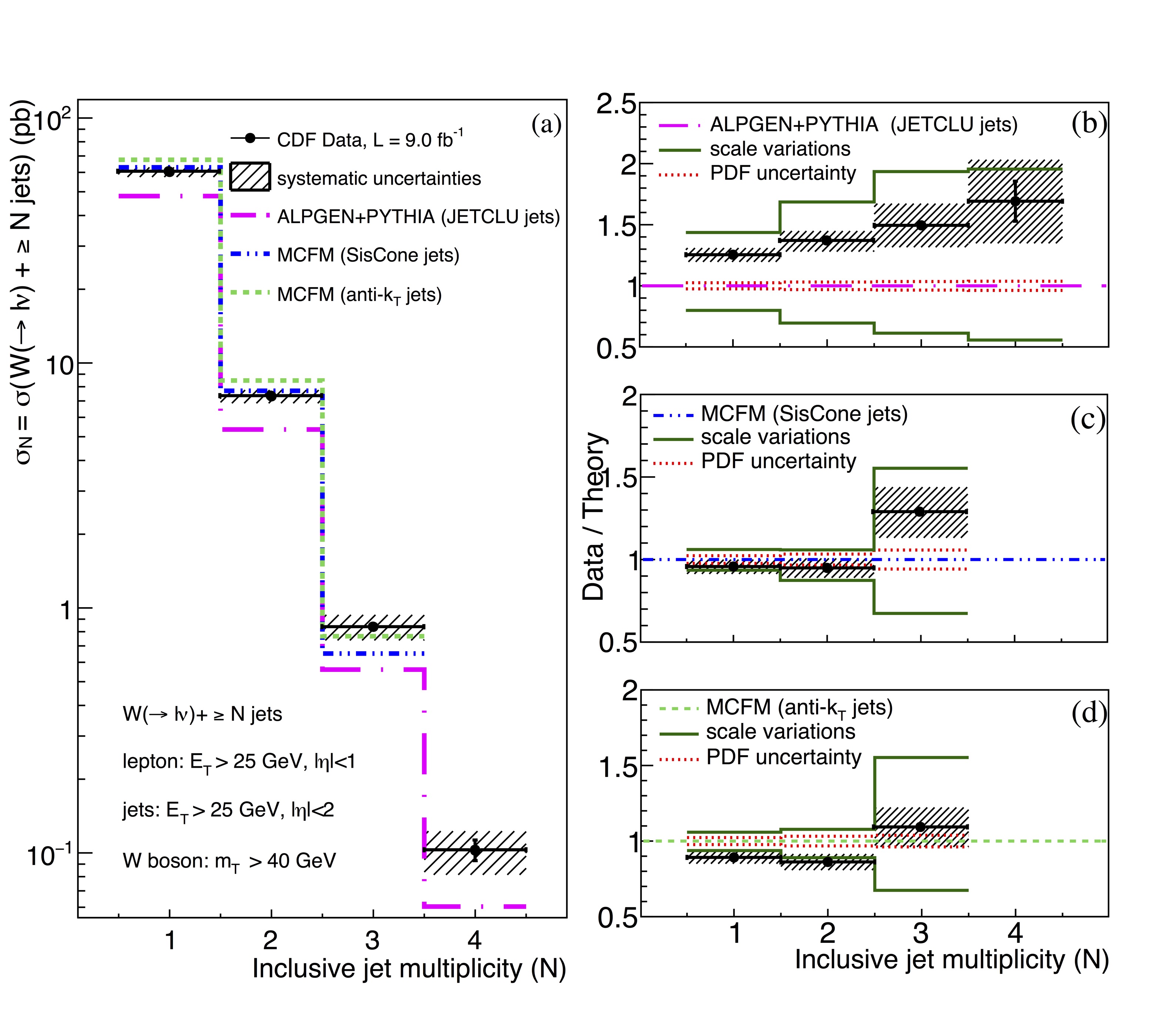}}\
	\end{center}	
	\caption{Measured inclusive jet cross sections (black dots), ($\sigma_N=  \sigma(W(\rightarrow \ell \nu) + \geqslant N\,\textrm{jets})$), as functions of the inclusive jet multiplicity for $W + \geqslant N$ jet events compared to the theoretical predictions described in Sec.~\ref{sec:VIII}.
	The panels show (a) the absolute comparisons, the ratios of the measured cross sections to (b) the \textsc{Alpgen+Pythia}  predictions and to (c) the \textsc{Mcfm} theoretical predictions corrected for npQCD prediction using SisCone and to (d) anti-$k_\textrm{T}$ jets. The shaded bands show the total systematic uncertainties, except for the 6\% luminosity uncertainty. The dashed and solid lines indicate the PDF uncertainties and the uncertainties corresponding to the variation of the factorization and renormalization scale $\mu$, respectively. \label{resultsI}}
\end{figure*}

\begin{figure*}[ht!]	
 \captionsetup{labelsep=period, format=plain, justification=justified}	
 \begin{center}	
{\includegraphics[width=\textwidth]{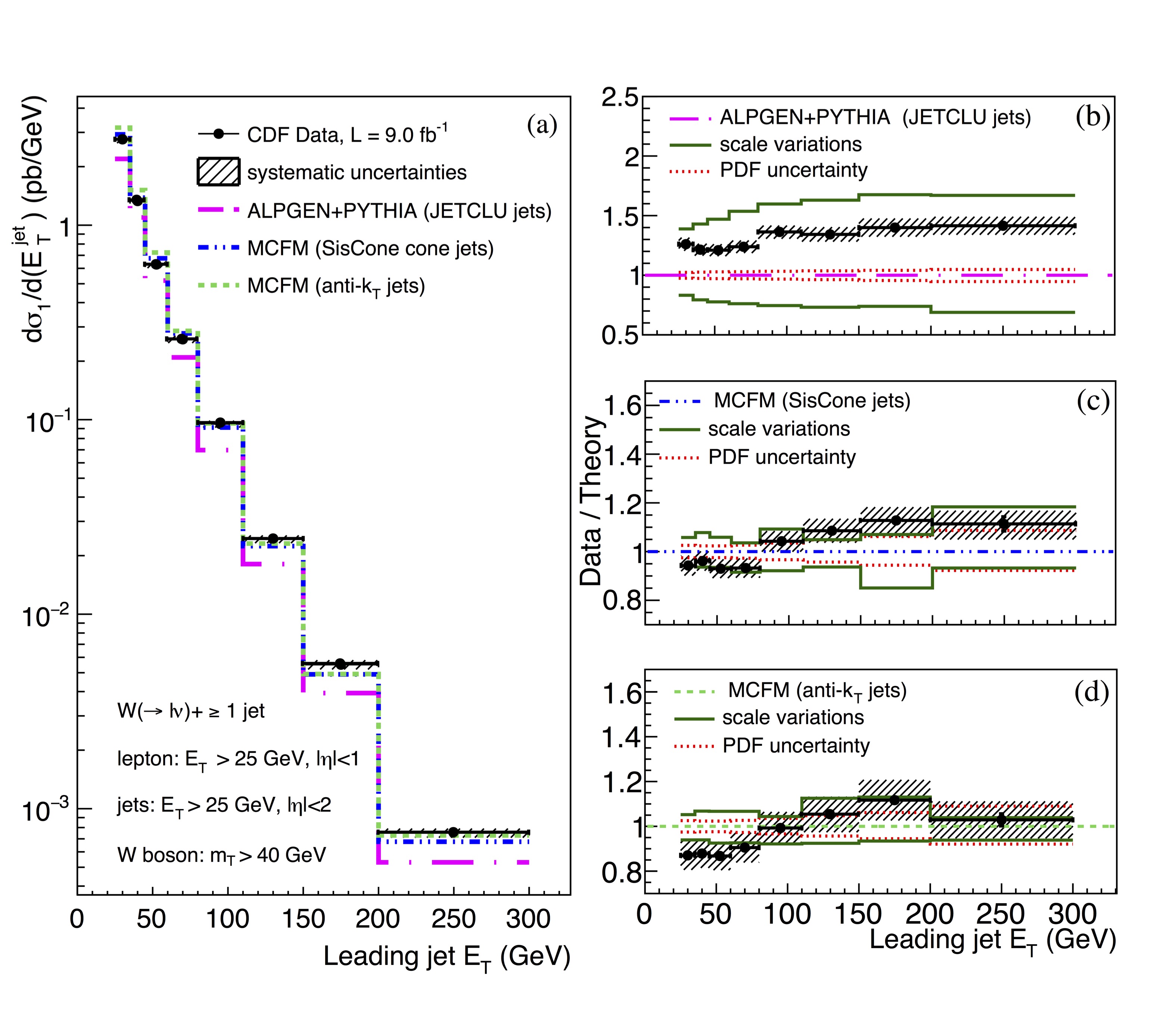}}\
	\end{center}	
	\caption{Measured differential cross sections, ($d\sigma_{1}/dE^{\rm jet}_\textrm{T}$) , as functions of the leading jet $E_\textrm{T}$ for $W + \geqslant 1$ jet events compared to the theoretical predictions described in Sec.~\ref{sec:VIII}. The panels show (a) the absolute comparisons, the ratios of the measured cross sections to (b) the \textsc{Alpgen+Pythia}  predictions and to (c) the \textsc{Mcfm} theoretical predictions corrected for npQCD prediction using  SisCone and  to (d) anti-$k_\textrm{T}$ jets. The shaded bands show the total systematic uncertainties, except for the 6\% luminosity uncertainty. The dashed and solid lines indicate the PDF uncertainties and the uncertainties corresponding to the variation of the factorization and renormalization scale $\mu$, respectively. \label{resultsII}}
\end{figure*}

The cross sections as functions of the jet multiplicity ($\sigma_N=  \sigma(W(\rightarrow \ell \nu) + \geqslant N\,\textrm{jets})$) and the leading-jet $E_\textrm{T}$ ($d\sigma_{1}/dE^{\rm jet}_\textrm{T}$) are shown in Figs.\,\ref{resultsI} and \ref{resultsII}. The \textsc{Alpgen+Pythia} and the \textsc{Mcfm} predictions, corrected for npQCD effects, are included for comparison. Both the measurements and the predictions are particle-level cross sections restricted to the following requirements on the final-state particles: only one central ($|\eta|<$1) lepton with $E_\textrm{T}>25$\,GeV and at least one jet with $E_\textrm{T}>25$\,GeV and pseudorapidity $|\eta|<$2. The reconstructed transverse mass of the $W$ boson is required to be greater than 40\;GeV$/c^{\,2}$. Jets are reconstructed using the JETCLU algorithm with a radius parameter of 0.4 in the measurement and in the \textsc{Alpgen+Pythia} predictions while for the \textsc{Mcfm} predictions, the algorithms used are the anti-$k_\textrm{T}$ and cone algorithm. 

Panels (b)--(d) in Figs.\,\ref{resultsI} and \ref{resultsII} show that the \textsc{Alpgen+Pythia} predictions are affected by a large uncertainty due to variations of the renormalization and factorization scales. The data are consistent with the predictions within these uncertainties. 
The NLO \textsc{Mcfm} predictions corrected for npQCD effects agree with the measurements despite the differences in the jet-reconstruction algorithms used in \textsc{Mcfm} and this analysis. This observation would be more significant if similar infrared-safe reconstruction algorithms had been used for both the theory and the data analysis. However, as observed in Ref.~\cite{jetography}, the differences between the  SisCone and anti-$k_\textrm{T}$ algorithms used for the \textsc{Mcfm} NLO predictions is indicative of the bias introduced by the use of JETCLU. These differences are observed to be smaller than the uncertainties in the measurements. Moreover, similar agreement between data and theory has been observed previously in analogous comparisons~\cite{W+3jets}.

\begin{figure*}[ht!]	
 \captionsetup{labelsep=period, format=plain, justification=justified}	
 \begin{center}	
{\includegraphics[width=0.65\textwidth]{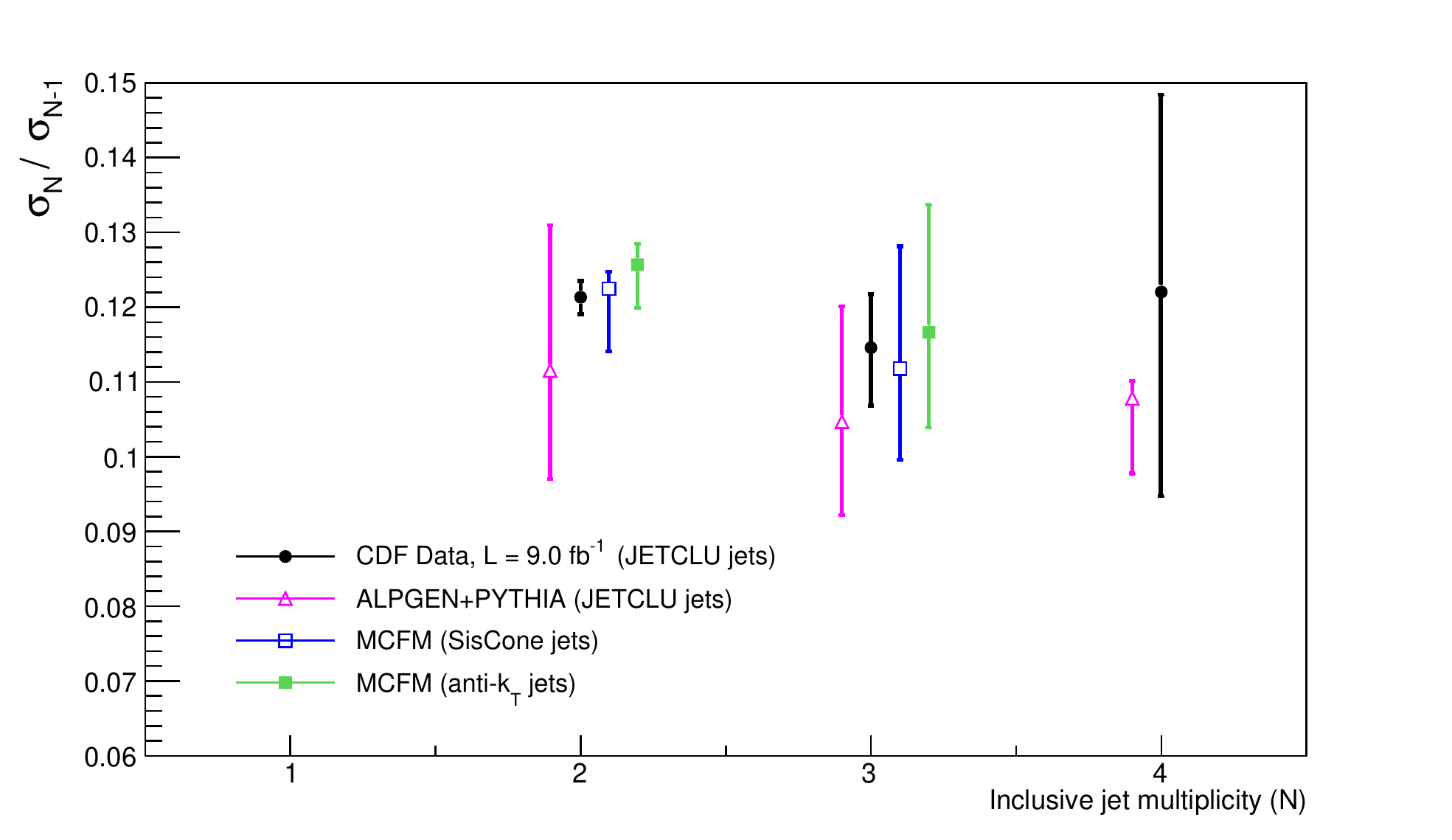}}\
	\end{center}	
	\caption{Ratio between inclusive-jet differential cross sections ($\sigma_{N}/\sigma_{N-1}$ for $N=2, 3, 4$) in data (black dots),  \textsc{Alpgen+Pythia} predictions (triangles) and \textsc{Mcfm} predictions corrected for npQCD effects using SisCone cone (open squares) and anti-$k_\textrm{T}$ jets (solid squares). The $\sigma_{N}/\sigma_{N-1}$ \textsc{Mcfm} prediction for $N=3$ is a LO ratio, while no \textsc{Mcfm} is available for for $N>3$. Error bars show the total uncertainties. \label{resultsIII}}
\end{figure*}

Figure \ref{resultsIII} shows the ratios between inclusive jet multiplicity cross sections ($\sigma_{N}/\sigma_{N-1}$). The \textsc{Mcfm} prediction for $N=3$ is calculated using a LO prediction for both the numerator ($\sigma_3$) and the denominator ($\sigma_2$), while NLO predictions are used for the numerator and the denominator  for $N=2$. The theoretical predictions agree with the measurements. The cross section ratios are sensitive to the strong-interaction coupling, and display no discernible dependence on the jet multiplicity, as expected.

\section{\label{sec:X} Summary and Conclusions}

Measurements of differential inclusive cross sections for the production of jets in association with a $W$ boson, using $9.0$\;fb$^{-1}$ of $p\bar{p}$ collision data collected by the CDF experiment at the Tevatron, are reported. The differential cross sections as functions of jet multiplicity and leading-jet $E_\textrm{T}$ are measured independently for the $W(\rightarrow e \nu)$ and $W(\rightarrow \mu \nu)$ decay modes and are combined at the particle level after unfolding detector acceptance and resolution effects.  Measurements are performed in the kinematic region $E_\textrm{T}^{\ell}>25$\,GeV, $|\eta^{\ell}| < 1$, $E_\textrm{T}^{\rm jet}>25$\,GeV, $|\eta^{\rm jet}| < 2$ and $m_\textrm{T}^W>40$\,GeV/$c^2$ and jets are reconstructed with the JETCLU algorithm. 
Cross sections are compared with the theoretical predictions of the \textsc{Alpgen} generator interfaced with \textsc{Pythia} (enhanced leading-order QCD predictions) and the \textsc{Mcfm} generator (next-to-leading order QCD predictions) corrected for nonperturbative QCD effects. The theoretical predictions are mainly affected by the uncertainty on  the factorization and renormalization scale. This uncertainty of the \textsc{Alpgen+Pythia} predictions is significantly larger than the uncertainty on the measurements, whereas for the \textsc{Mcfm} predictions it is comparable to the experimental uncertainty. 
The agreement with these predictions observed for the measurements reported here suggests that the NLO perturbative QCD calculations properly model the jet multiplicity and jet $E_T$ distributions of the $W+$jets process. 
The ratio of the lepton coupling constants reported in Table~\ref{U} is consistent with lepton universality and validates the procedure used to evaluate the QCD background.
The production of a $W$ boson in association with jets is among the dominant backgrounds in current measurements and searches for non-standard-model physics at the Large Hadron Collider. The proper modeling of this process, supported by our work, is therefore important to consolidate and enhance the physics reach of Large Hadron Collider studies.

\begin{acknowledgments}
We thank the Fermilab staff and the technical staffs of the participating institutions for their vital contributions. This work was supported by the U.S. Department of Energy and National Science Foundation; the Italian Istituto Nazionale di Fisica Nucleare; the Ministry of Education, Culture, Sports, Science and Technology of Japan; the Natural Sciences and Engineering Research Council of Canada; the National Science Council of the Republic of China; the Swiss National Science Foundation; the A.P. Sloan Foundation; the Bundesministerium f\"ur Bildung und Forschung, Germany; the Korean World Class University Program, the National Research Foundation of Korea; the Science and Technology Facilities Council and the Royal Society, UK; the Russian Foundation for Basic Research; the Ministerio de Ciencia e Innovaci\'{o}n, and Programa Consolider-Ingenio 2010, Spain; the Slovak R\&D Agency; the Academy of Finland; the Australian Research Council (ARC); and the EU community Marie Curie Fellowship contract 302103.
\end{acknowledgments}

\begin{appendix}

\section{\label{APPI} Background validation plots}

\begin{figure*}[ht!]
 \captionsetup{labelsep=period, format=plain, justification=justified}	
	\begin{center}
	\vspace{-0.0cm}
\subfloat[]{\includegraphics[width=0.45\textwidth]{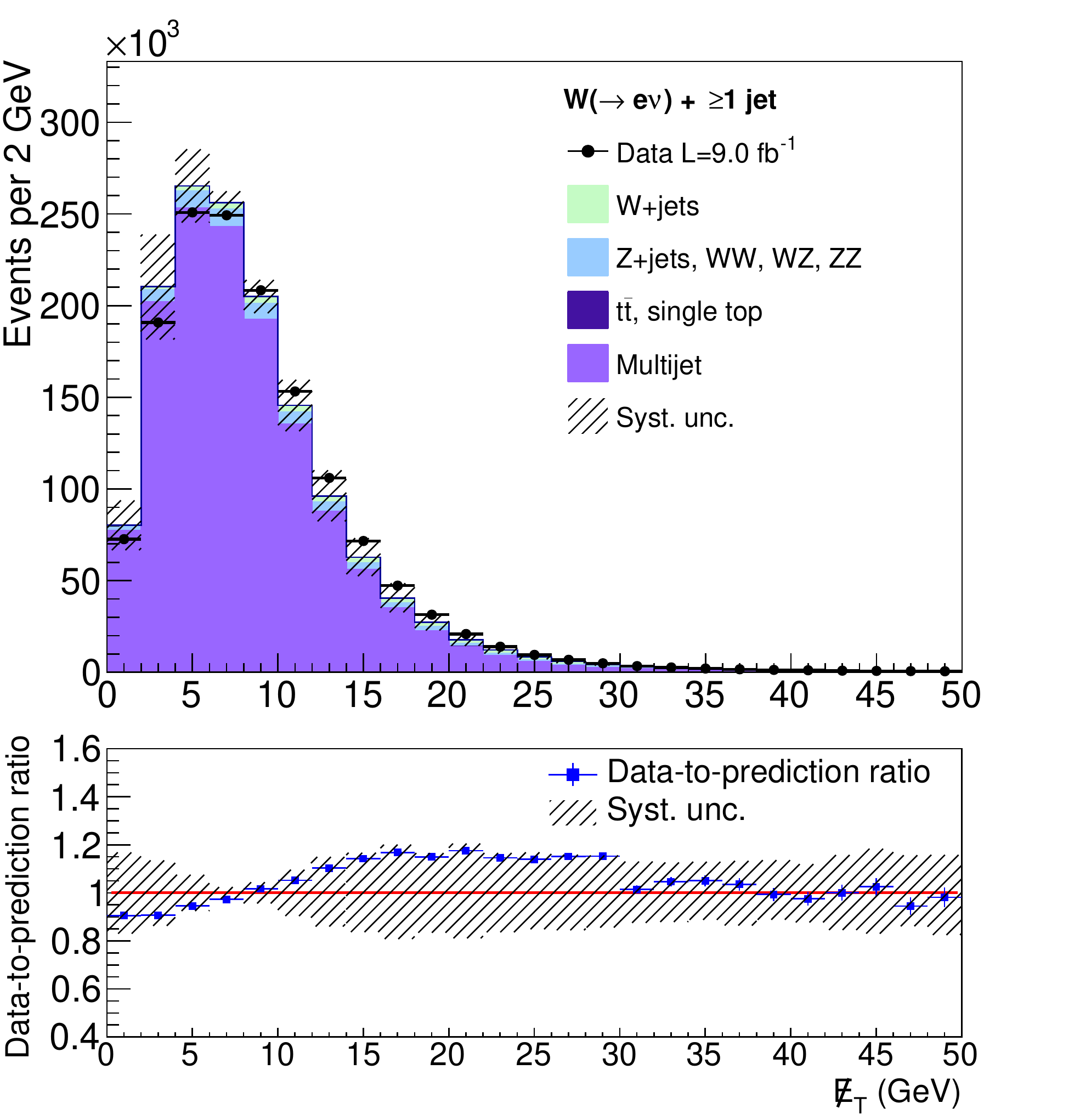}}
\subfloat[]{\includegraphics[width=0.45\textwidth]{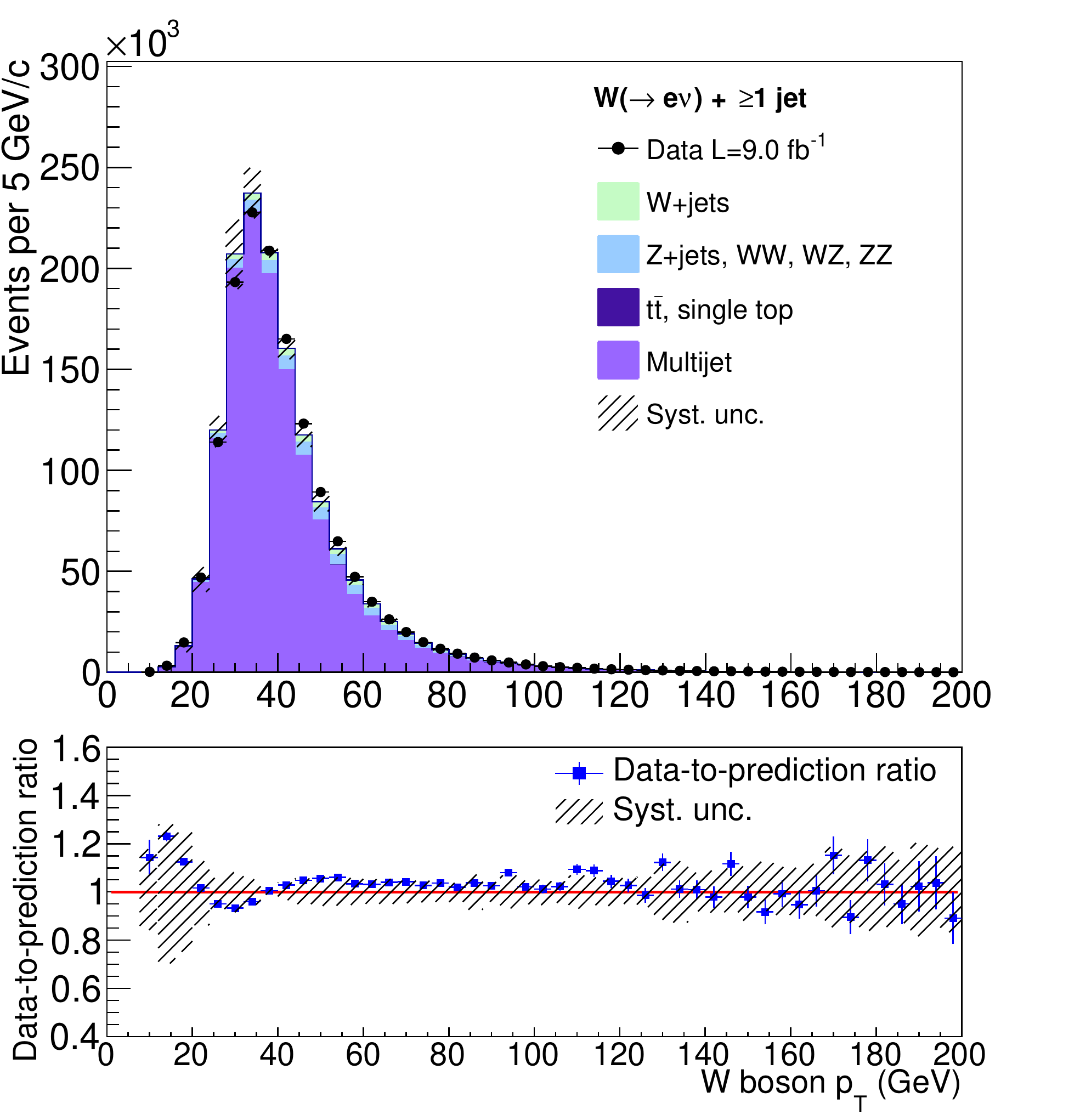}}\\	
\vspace{-0.0cm}
\subfloat[]{\includegraphics[width=0.45\textwidth]{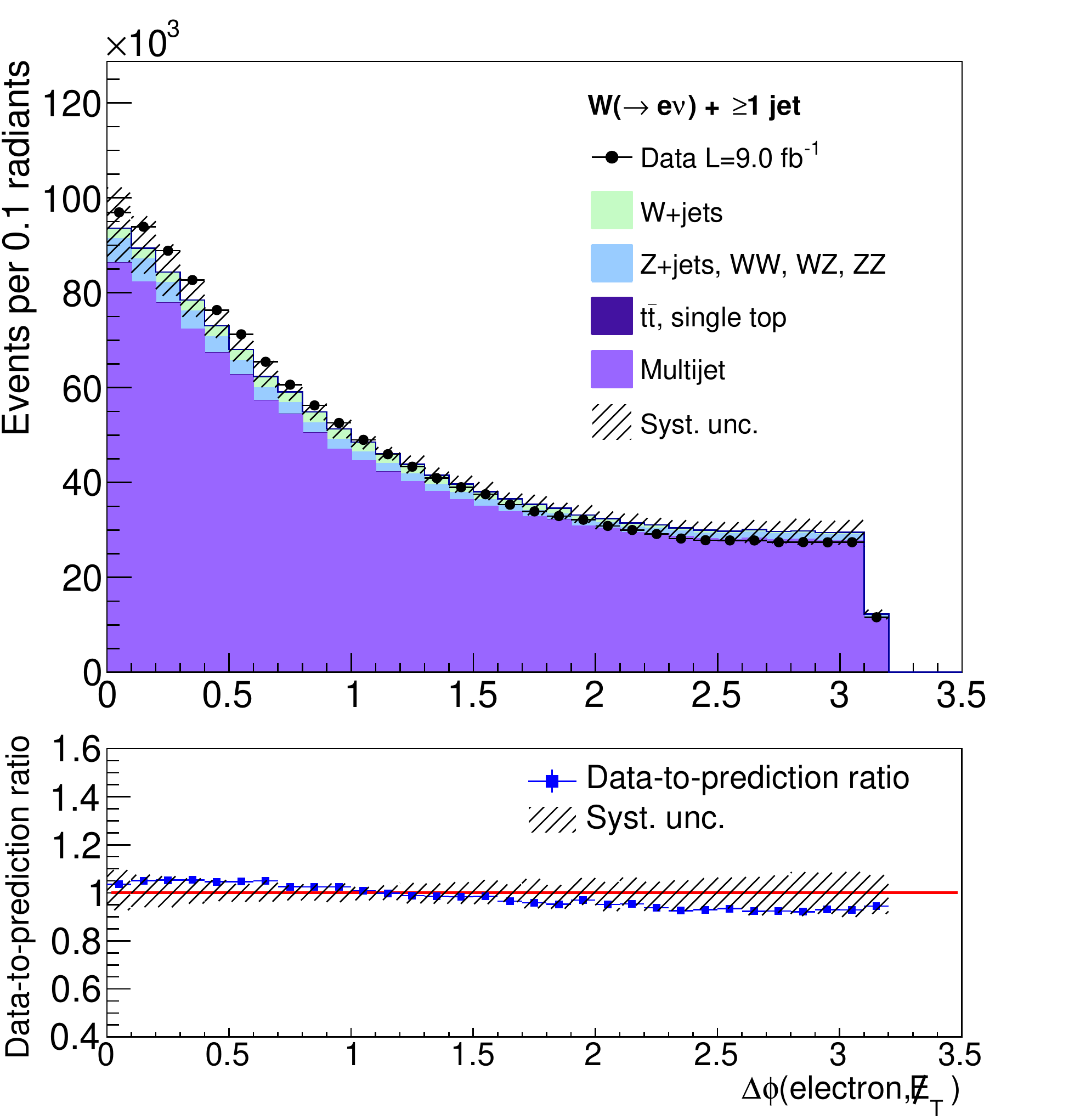}}
\subfloat[]{\includegraphics[width=0.45\textwidth]{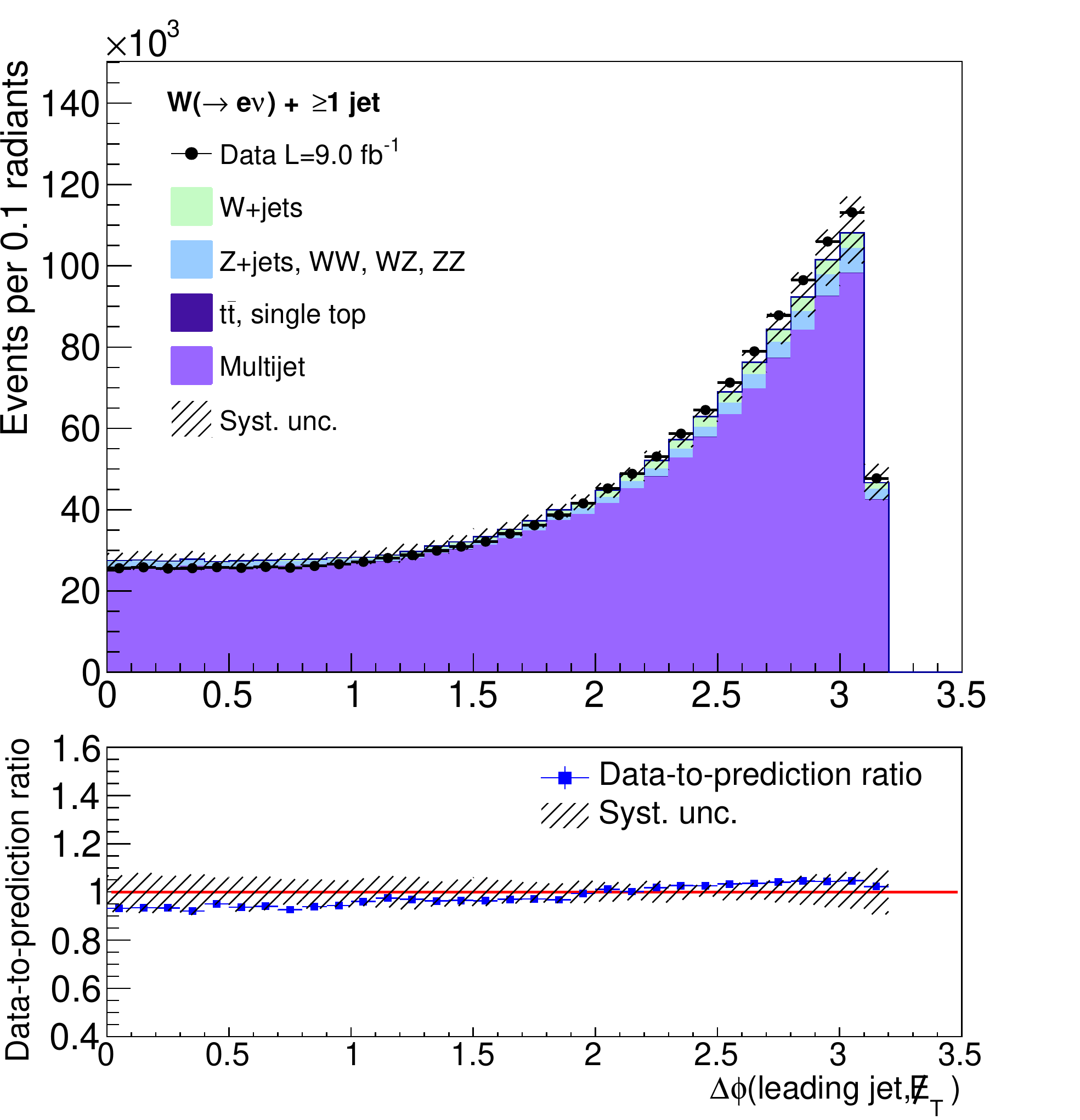}}\\	
\end{center}	
\vspace{-0.0cm}	
 \caption{Validation plots of the background model in the electron channel: comparison between the data and the prediction for the distributions of (a) missing transverse energy $\met$, (b) $W$-boson $p_\textrm{T}$, (c) azimuthal distance between the electron and the $\vecmet$ ($\Delta \phi (e,\vecmet)$) and (d) azimuthal distance between the leading jet and the $\vecmet$ ($\Delta \phi ({\rm leading\ jet},\vecmet)$) in the control region.  The data are represented by the black points while the signal and the background predictions are represented by stacked histograms. Systematic uncertainties on the predictions are indicated by the shaded areas (see Sec.~\ref{sec:IV} for discussion on systematic uncertainty). The lower plots show the ratios between the data and the corresponding predictions. \label{model_ele}}
\end{figure*}

\begin{figure*}[ht!]
 \captionsetup{labelsep=period, format=plain, justification=justified}	
	\begin{center}
	\vspace{-0.0cm}	
\subfloat[]{\includegraphics[width=0.45\textwidth]{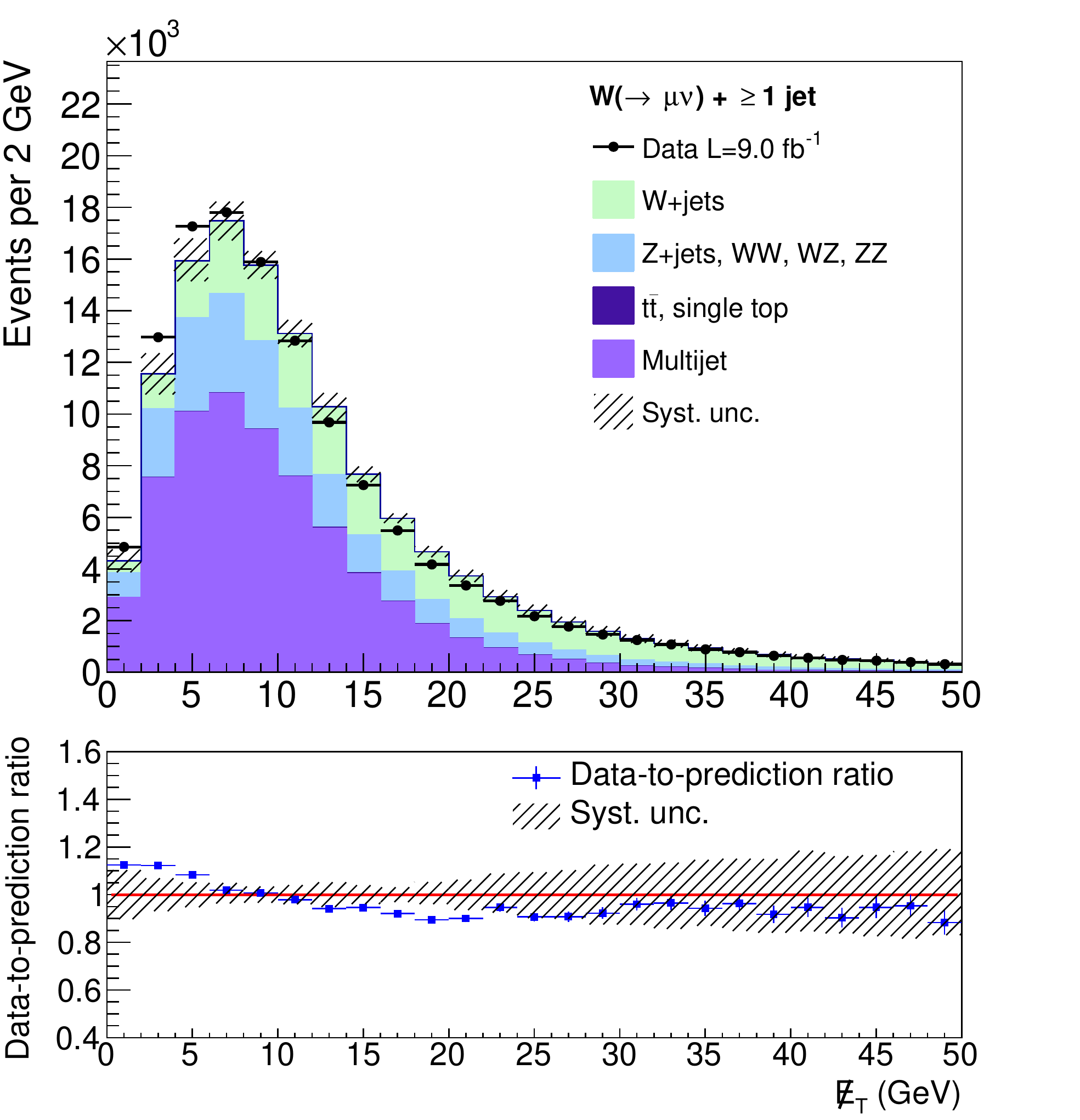}}
\subfloat[]{\includegraphics[width=0.45\textwidth]{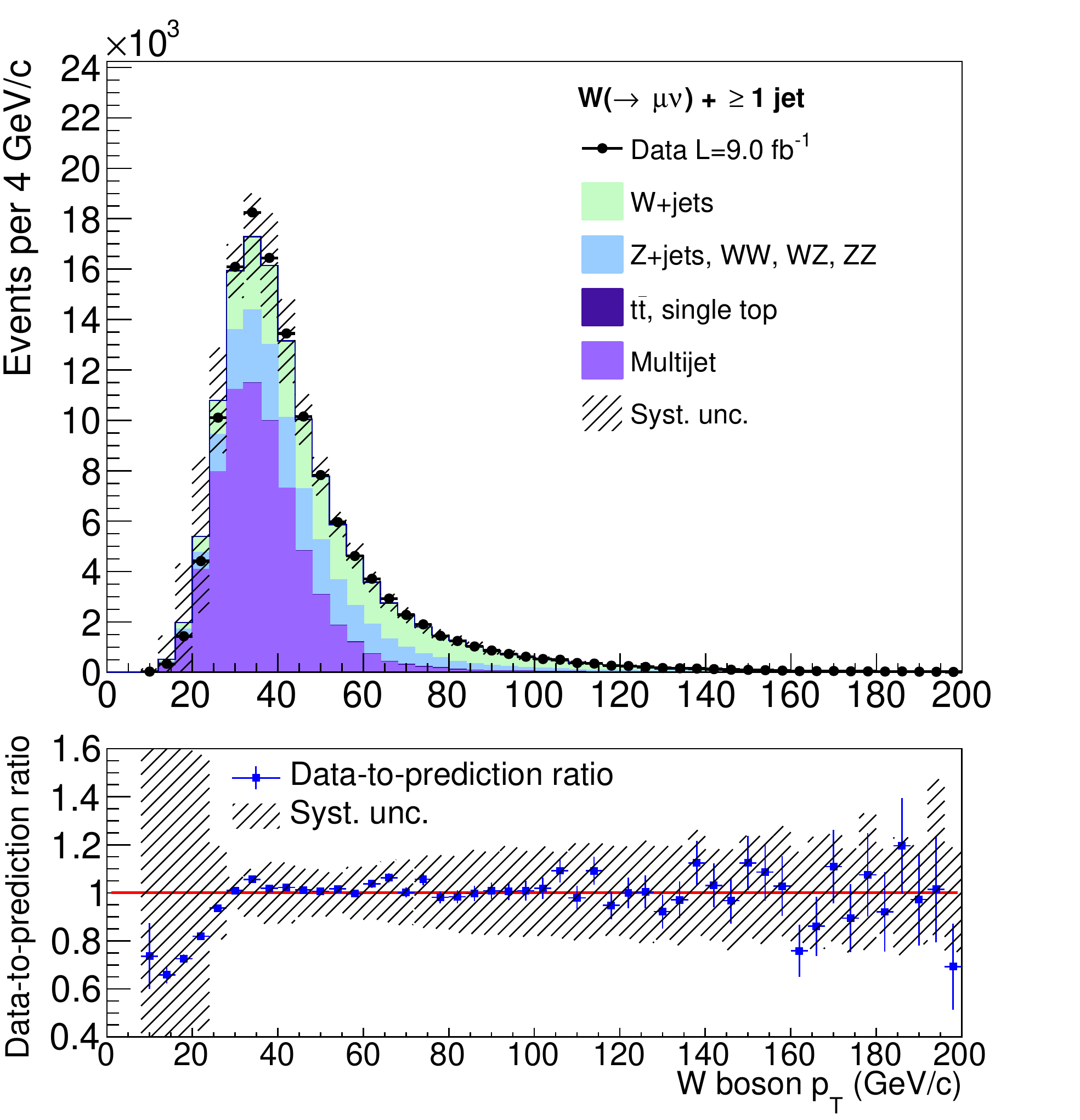}}\\	
\vspace{-0.1cm}
\subfloat[]{\includegraphics[width=0.45\textwidth]{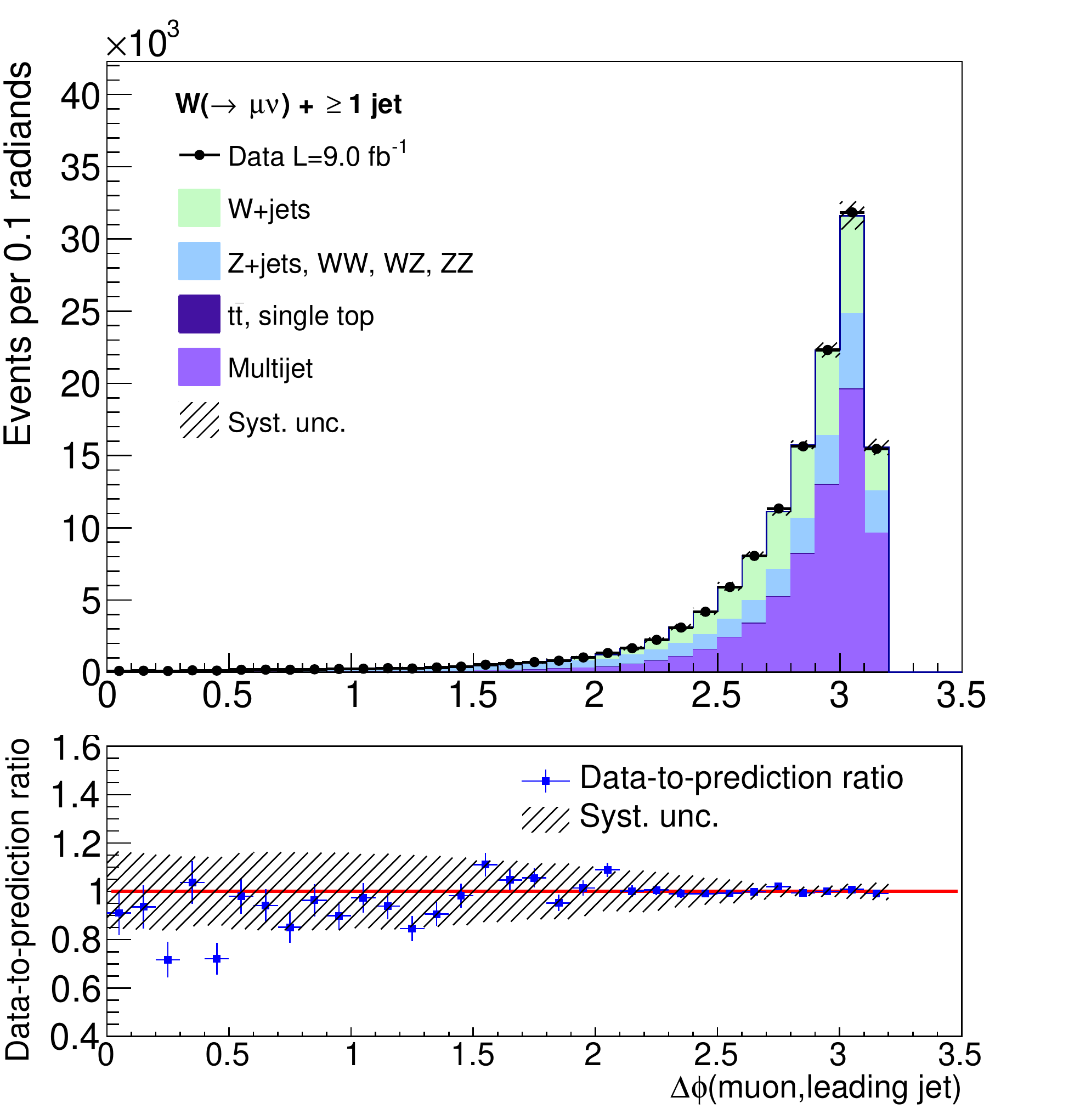}}
\subfloat[]{\includegraphics[width=0.45\textwidth]{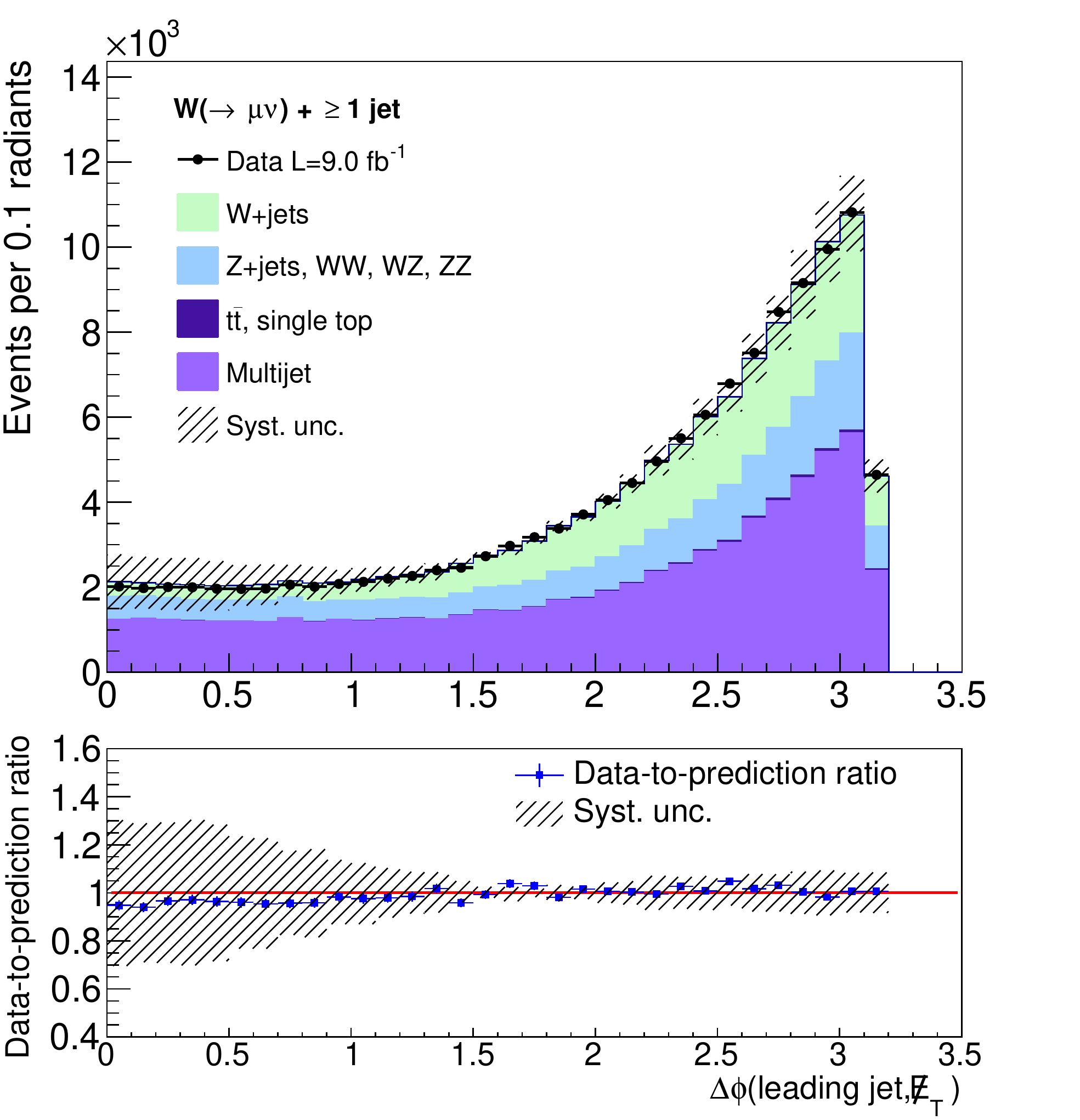}}\\	
\vspace{-0.0cm}
\end{center}		
\caption{Validation plots of the background model in the muon channel: comparison between the data and the prediction for the distributions of (a) missing transverse energy $\met$, (b) $W$-boson $p_\textrm{T}$, (c) azimuthal distance between the muon and the leading jet ($\Delta \phi (\mu,{\rm leading\ jet})$) and (d) azimuthal distance between the leading jet and the $\vecmet$ ($\Delta \phi ({\rm leading\ jet},\vecmet)$) in the control region. The data are represented by the black points while the signal and the background predictions are represented by stacked histograms. Systematic uncertainties on the predictions are indicated by the shaded areas (see Sec.~\ref{sec:IV} for discussion on systematic uncertainty). The lower plots show the ratios between the data and the corresponding predictions. \label{model_muo}}
\end{figure*}

As a consistency check of the background models, data and predictions are compared in a control region where the background contributions are expected to be much larger than the signal, as explained in Sec.~\ref{sec:IV}. In addition to those shown in Fig.~\ref{model} of Sec.~\ref{sec:IVC}, other examples of background-modeling validation plots for the electron and muon channels are shown in  Figs.~\ref{model_ele} and~\ref{model_muo}, respectively. The reasonably good agreement between the data and the predictions supports the assumption that the multijet model and the MC simulations adequately describe the contributions from the background processes.

\FloatBarrier

\section{\label{APPII} Detector response and acceptance matrices for the particle-level results}

\begin{figure}[ht!]
 \captionsetup{labelsep=period, format=plain, justification=justified}	
	\begin{center}
\subfloat[]{\includegraphics[width=0.45\textwidth]{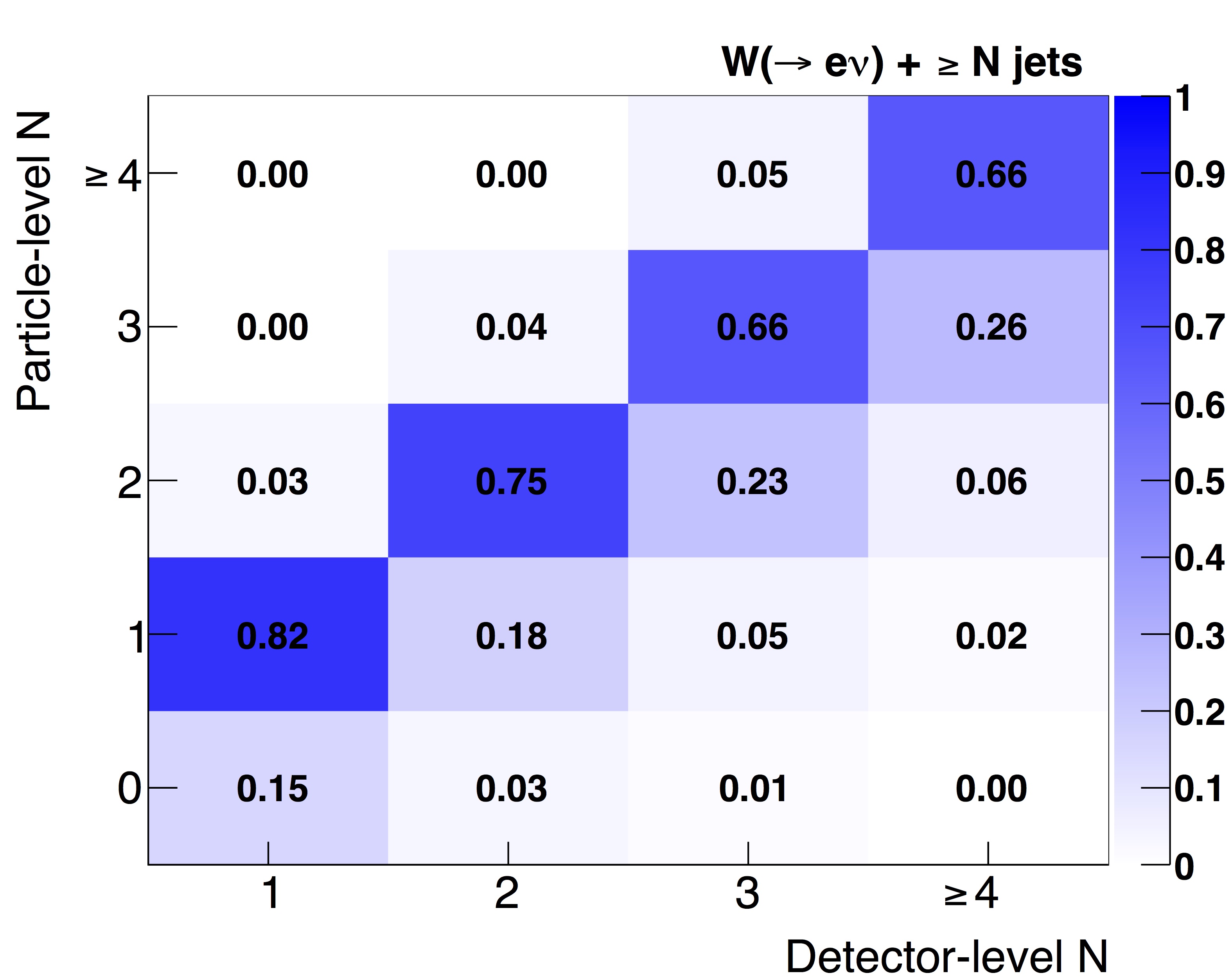}}\\
\subfloat[]{\includegraphics[width=0.45\textwidth]{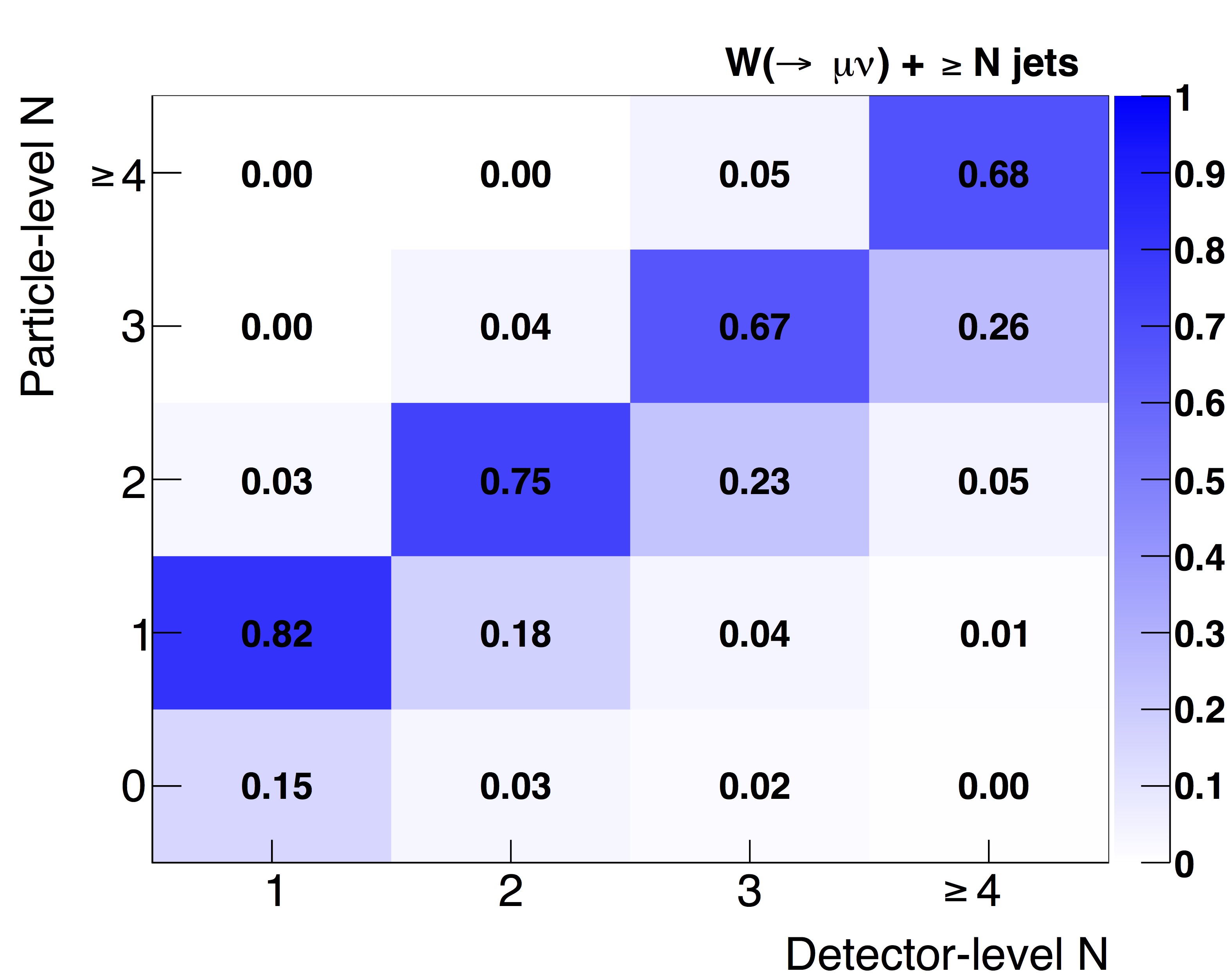}}\\	
\end{center}		
\caption{Detector response matrices that describe the migration between bins corrected by the unfolding procedure for the measurement of $\sigma(W(\rightarrow \ell \nu) + N\,\textrm{jets})$ (a) in the electron channel ($\ell = e$), and (b) in the muon channel ($\ell = \mu$). \label{fig:correlationsN}}
\end{figure}

\begin{figure}[ht!]	
	\begin{center}
\subfloat[]{\includegraphics[width=0.45\textwidth]{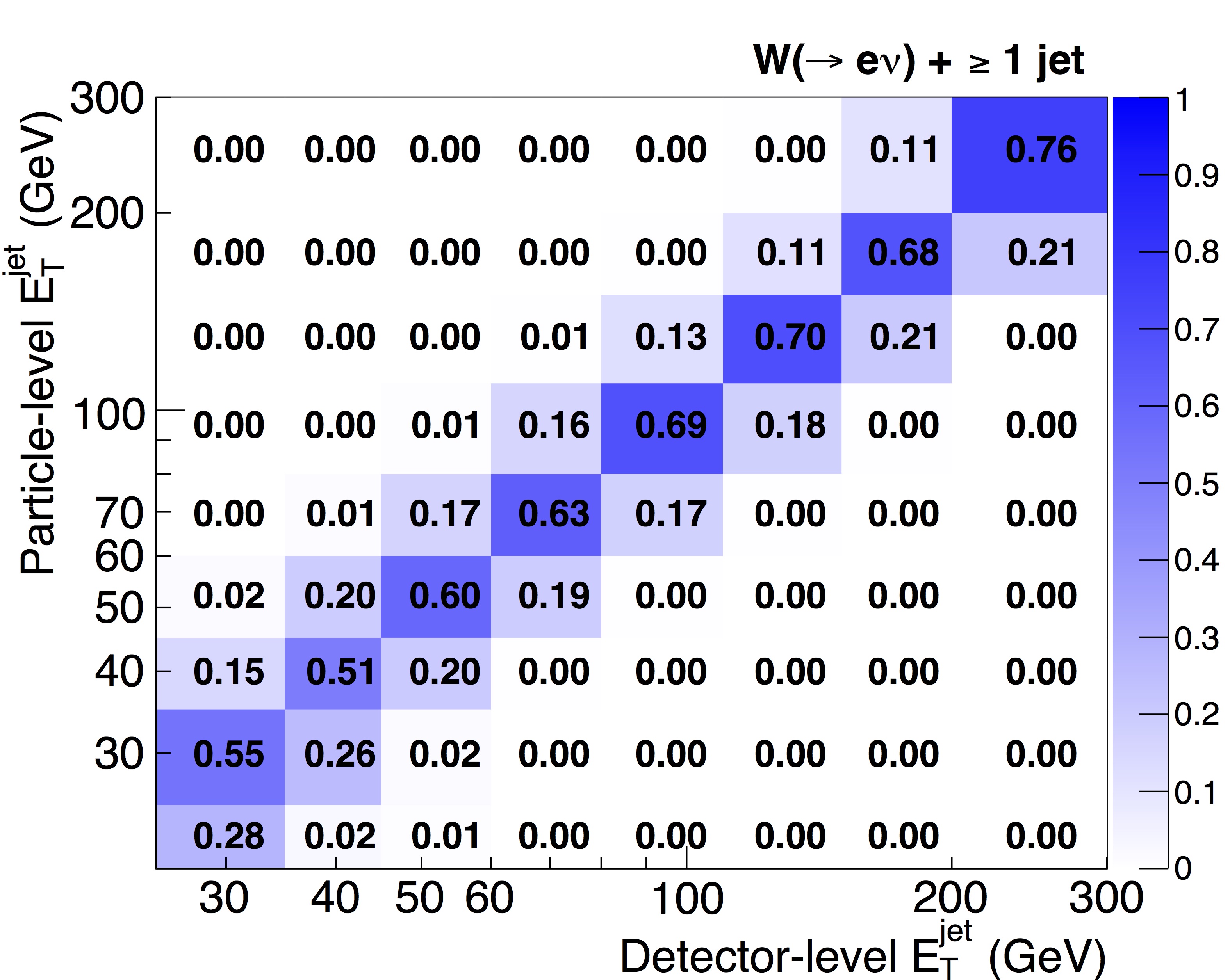}}\\
\subfloat[]{\includegraphics[width=0.45\textwidth]{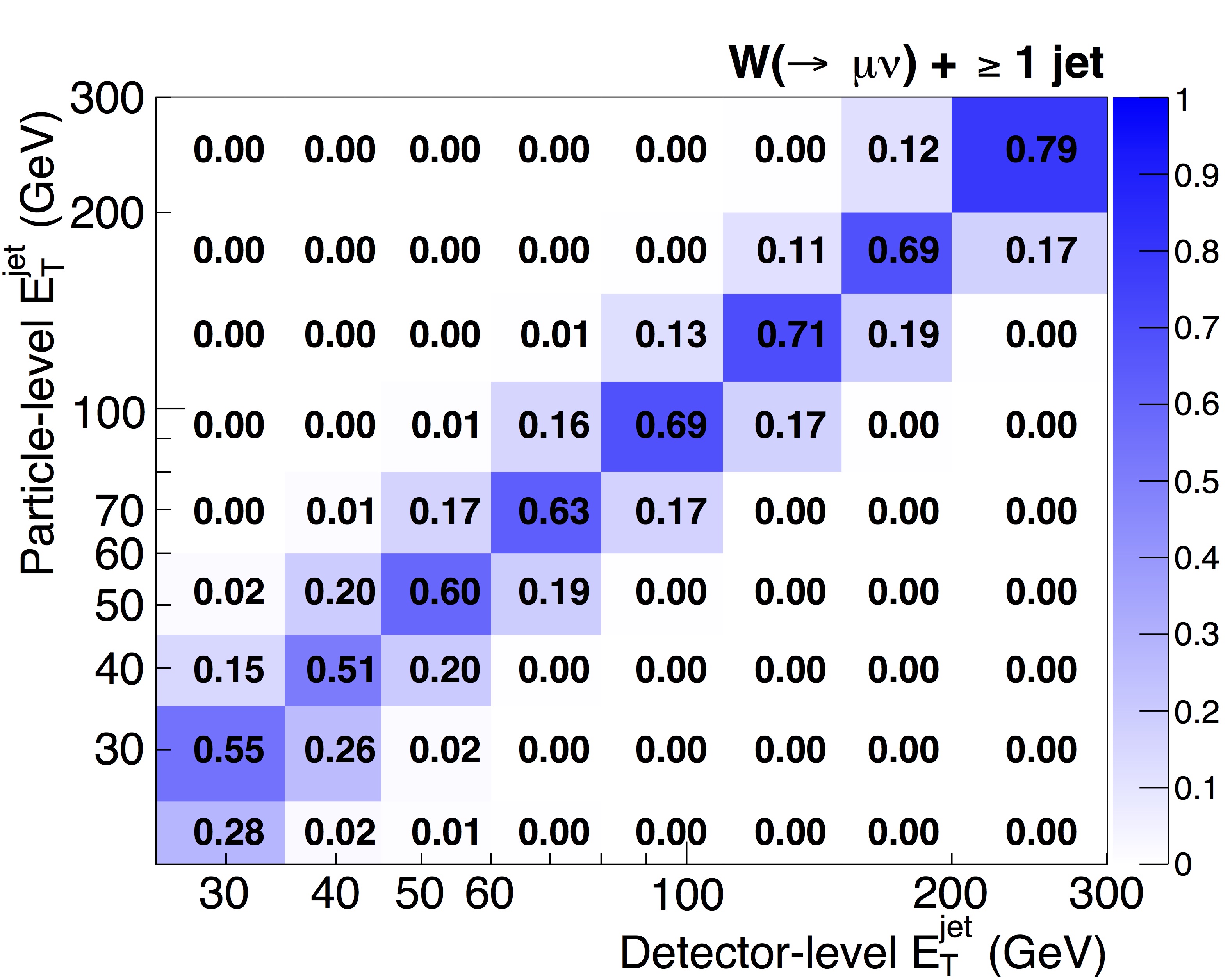}}\\	
\end{center}		
\caption{Detector response matrices that describe the migration between bins corrected by the unfolding procedure for the measurement of $d\sigma_{1}/dE^{\rm jet}_{\rm T}$, where $\sigma_1 = W(\rightarrow \ell \nu )+ \geqslant 1\,\textrm{jet}$ (a) for the electron ($\ell =e$), and (b) for the muon ($\ell = \mu$) channels. \label{fig:correlationsJ}}
\end{figure}

 The detector response matrices account for migrations of events between the bins in which the cross sections are measured and the corresponding bins at particle level. They are are non-diagonal matrices as shown in figures~\ref{fig:correlationsN} and~\ref{fig:correlationsJ} , where the fraction of detector-level events reconstructed from each particle-level bin is presented.
 
\FloatBarrier

The acceptance matrices quantify the probabilities of an event to be detected and selected, as functions of the particle-level quantities $N$~jets and $E_{\rm T}^{\rm jet}$. The acceptance matrices are diagonal, and the entries are listed in Tables~\ref{tab:AN}~and~\ref{tab:AJ}. About 35-50\% of the particle-level events events are selected in each bin. The uncertainties reported include only the statistical contribution.

\begin{table*}[ht!]
 \captionsetup{labelsep=period, format=plain, justification=justified}	
  \centering
   \caption{Diagonal elements of the acceptance matrices for the particle-level measurement of  $\sigma(W(\rightarrow e \nu) + N\,\textrm{jets})$ and $\sigma(W(\rightarrow \mu \nu) + N\,\textrm{jets})$. The uncertainties reported are only statistical. \label{tab:AN}}
  \begin{tabular*}{\textwidth}{@{\extracolsep{\fill} }lcc}
\hline
\hline
   &  $W(\rightarrow e \nu)$ channel & $W(\rightarrow \mu \nu)$ channel \\
   \hline
   $N=1$\,jet     & (35.33 $\pm$ 0.04)\% & (35.38 $\pm$ 0.04)\% \\
   $N=2$\,jets  & (36.9 $\pm$ 0.1)\%  & (37.0 $\pm$ 0.1)\% \\
   $N=3$\,jets & (37.4 $\pm$ 0.2)\% & (37.6 $\pm$ 0.2)\%\\
   $N\geqslant 4$\,jets  & (38.3 $\pm$ 0.3)\% & (38.2 $\pm$ 0.2)\%  \\
 \hline
  \hline
\end{tabular*} 
 \vspace{1.cm}
 \captionsetup{labelsep=period, format=plain, justification=justified}	
  \centering
   \caption{Diagonal elements of the acceptance matrices for the particle-level measurement of  $d\sigma_{1}/dE^{\rm jet}_{\rm T}$ (where $\sigma_1 = W(\rightarrow \ell \nu )+ \geqslant 1\,\textrm{jet}$) for the electron ($\ell =e$) and muon ($\ell = \mu$) channels. \label{tab:AJ}}
   \begin{tabular*}{\textwidth}{@{\extracolsep{\fill} }lcc}
\hline
\hline
&  $W(\rightarrow e \nu)$ channel & $W(\rightarrow \mu \nu)$ channel \\
\hline
25\,GeV $\leqslant E^{\textrm\ jet}_\textrm{T} < $ 35\,GeV & (35.03 $\pm$ 0.05)\% & (35.09 $\pm$ 0.05)\% \\
35\,GeV $\leqslant E^{\textrm\ jet}_\textrm{T} <  $ 45\,GeV & (35.06 $\pm$ 0.07)\% & (35.12 $\pm$ 0.07)\% \\
45\,GeV $\leqslant E^{\textrm\ jet}_\textrm{T} <  $ 60\,GeV & (35.58 $\pm$ 0.08)\% & (35.5 $\pm$ 0.08)\% \\
60\,GeV $\leqslant E^{\textrm\ jet}_\textrm{T} <  $ 80\,GeV & (36.2 $\pm$ 0.1)\% & (36.4 $\pm$ 0.1)\% \\
80\,GeV $\leqslant E^{\textrm\ jet}_\textrm{T} <  $ 110\,GeV & (38.5 $\pm$ 0.2)\% & (38.4 $\pm$ 0.2)\% \\
110\,GeV $\leqslant E^{\textrm\ jet}_\textrm{T} <  $ 150\,GeV & (41.7 $\pm$ 0.3)\% & (41.7 $\pm$ 0.3)\% \\
150\,GeV $\leqslant E^{\textrm\ jet}_\textrm{T} <  $ 200\,GeV & (45.4 $\pm$ 0.6)\% & (45.5 $\pm$ 0.6)\% \\
200\,GeV $\leqslant E^{\textrm\ jet}_\textrm{T} <  $ 300\,GeV & (49.5 $\pm$ 1.3)\% & (51.4 $\pm$ 1.3)\% \\
 \hline
  \hline
\end{tabular*} 
 \vspace{1.cm}
 \end{table*}

\end{appendix}


\begin{thebibliography}{99}
\bibitem{transverse}
Transverse energy (momentum) is the energy (momentum ) projected into the plane transverse to the beam direction; a detailed explanation is given in Sec.~\ref{sec:II}.
\bibitem{vertex} 
The primary vertex is reconstructed by fitting a subset of good-quality tracks. If there are multiple possible vertices, the primary vertex is chosen to be the one associated with the charged particle with the highest $p_\textrm{T}$.
\bibitem{isolation}  
Isolation is defined as the ratio of the calorimeter transverse energy in a cone of opening angle $\Delta R\equiv \sqrt{(\Delta\eta)^2+(\Delta\phi)^2}=0.4$ around the lepton (not including that of the lepton energy itself) divided by the lepton $E_\textrm{T}$ (for the electron) or $p_\textrm{T}$ (for the muon). An electron or a muon candidate is defined as isolated when the isolation variable is less than 0.1.
\bibitem{MTW}
\sloppy{The transverse mass of the $W$ candidate is defined as $m_\textrm{T}^{W}= \sqrt{2p_\textrm{T}^{\ell}\met[1-\cos(\Delta\phi(\ell,\vecmet))]}$,  where $p_\textrm{T}^{\ell}$ is the momentum vector of the lepton (electron or muon: $\ell=e, \mu$), $\vecmet$ is the missing transverse energy and $\Delta\phi(\ell,\vecmet)$ is the azimuthal separation between the lepton momentum vector (${\ell}$) and the $\vecmet$. The missing transverse energy vector ($\vecmet$) is defined as minus the vector sum of the energy depositions in the calorimeter cells, projected in the transverse plane. The jet energy corrections and a correction for the muon energy depositions are included in the definition. In this paper the missing transverse energy ($\met$) is referred as the magnitude of the missing transverse energy vector.}
\bibitem{UE} 
Additional interactions between the partons not involved in the primary interaction are not negligible, and their contribution to the final state of the hard-scattering process is known as the ``underlying event''.
\bibitem{QG} 
Studies conducted in Ref.~\cite{PRD89092001(2014)} have shown that the response to particle
showers that originate from gluons and those that originate originate from quarks are different.
\bibitem{IRCS} 
Infrared safety and collinear safety are properties of jet finding algorithms. An ``infrared safe'' jet-finding algorithm is an algorithm for which the  number of reconstructed jets is not sensitive to the presence of soft partons in the event, while  a ``collinear-safe'' jet-finding algorithm is an algorithm that constructs a single jet out of multiple particles that are collinear.
\bibitem{impactpar} 
The impact parameter is defined as the distance of closest approach in the $r-\phi$ plane between the particle and the incident beam.
\end{thebibliography}
\end{document}